\pdfoutput=1
\documentclass[aps,prx,twocolumn,superscriptaddress]{revtex4}

\usepackage{tabularx}
\usepackage{graphicx}% Include figure files
\usepackage{dcolumn}% Align table columns on decimal point
\usepackage{amsfonts,amsmath,amssymb,bm,mathrsfs}
\usepackage{subcaption}
\usepackage{xcolor}
\usepackage{float}
\usepackage[utf8]{inputenc}
\usepackage[colorlinks=true,allcolors=blue]{hyperref}
\usepackage{overpic}

\usepackage{cleveref}
\crefname{figure}{fig.}{figs.}
\Crefname{figure}{Fig.}{Figs.}
\Crefname{Table}{Table}{Tables}
\renewcommand{\thetable}{\arabic{table}}
\begin{document}
\title{Photocatalytic Water Splitting in GeC/ReS$_2$ van der Waals Heterostructures }
%Using Ehrenfest Electron-Ion Dynamics  }
\date{\today}

\author{Parisa Beyranvand}
\affiliation{Department of Physics, Isfahan University of Technology, Isfahan, 84156-83111, Iran}

\author{Ismaeil Abdolhosseini~Sarsari}\email{abdolhosseini@iut.ac.ir}
\affiliation{Department of Physics, Isfahan University of Technology, Isfahan, 84156-83111, Iran}

\author{Noejung Park}
\affiliation{Department of Physics, Ulsan National Institute of Science and Technology, Ulsan, 689-798, Republic of Korea}

\begin{abstract}
Photocatalytic water splitting is considered as one of the most promising approaches for the green hydrogen production. But the rapid recombination of photogenerated electron-hole pairs severely inhibits its photocatalytic efficiency. Construction of van der Waals heterostructures is a good way to suppress this recombination. However, in staggered band heterostructures, the charge transfer could occur either through the type-II or the direct Z-scheme mechanism. Therefore it is very important to identify which mechanism is dominant. In the present work, we systematically study the GeC/ReS$_2$ heterostructure using first-principles density functional theory (DFT). The system exhibits good thermodynamic, dynamic and thermal stability and an indirect band gap of 1.38 eV which greatly enhances the visible-light absorption. We find that the direct Z-scheme mechanism is the dominant interfacial charge transfer by analysing band alignment and Gibbs free energy pathways. The oxygen evolution reaction (OER) can be driven spontaneously in the pH range $4 \le \text{pH} \le 14$ under light irradiation for both type-II and Z-scheme pathways but the energy barrier is much lower for the Z-scheme mechanism. Moreover, the type-II mechanism is not suitable for hydrogen evolution reaction (HER), but the Z-scheme mechanism can promote HER well in the entire pH range ($0 \le \text{pH} \le 14$). Remarkably, the strain-free (pristine) condition exhibits a solar-to-hydrogen (STH) efficiency of 41.40\%, well above the 10\% threshold for commercial viability. The STH efficiency is further enhanced to 60.68\% under a tensile strain of +6\%. To summarise, the Z-scheme strain-tunable GeC/ReS$_2$ heterostructure is a promising and efficient photocatalyst for overall water splitting.
\end{abstract}

\maketitle

\section{Introduction}\label{sec:intro}
The development of renewable energy technologies is essential for meeting the growing global demand for clean and sustainable power. Among various renewable energy strategies, photocatalytic water splitting has attracted considerable attention, as it enables the direct conversion of solar energy into hydrogen fuel~\cite{Ra2014,Wang_2022}. Despite significant progress in the development of single-component photocatalysts, many systems
continue to suffer from fundamental limitations, including low quantum efficiency, rapid charge-carrier recombination, and insufficient light absorption~\cite{Singh2015}. To overcome these challenges and achieve efficient separation of photogenerated charge carriers, several strategies have been proposed, among which the construction of van der Waals heterostructures based on intrinsic two-dimensional materials has emerged as a particularly promising approach. Owing to their large specific surface area and short charge-transfer pathways, two-dimensional heterostructures offer favourable conditions for enhancing photocatalytic performance~\cite{Wang2021,Zhao2021,Amin2015,Ahmad2023}. These two-dimensional structures facilitate the effective separation of electrons and holes by creating an electric field at the interface region.

Among various heterostructure configurations, type-II and direct Z-scheme architectures are particularly intriguing. In a type-II heterostructure, the band-edge offset at the
interface drives spontaneous transfer of photogenerated electrons and holes into different constituent layers. Although this band alignment can enhance charge
separation, it often leads to carrier accumulation at lower energy levels, resulting in a substantial reduction of the overall redox power~\cite{Ghosh2024,Su2018}. In contrast, the direct Z-scheme heterostructures, inspired by natural photosynthesis, follow a fundamentally different charge-transfer mechanism. In these systems, recombination occurs selectively between electrons in the conduction band of one component and holes in the valence band of the other. Meanwhile, high-energy electrons and holes are retained in their respective bands.
This mechanism not only promotes effective charge separation but also preserves the high redox potential of the charge carriers. Consequently, the coexistence of efficient carrier separation and strong redox power constitutes a key advantage of direct Z-scheme heterostructures over conventional type-II models~\cite{Xu2021,Liu2019,Daming2021,Shi2022}.

Although type-II and direct Z-scheme heterostructures possess similar band edge alignments, their interfacial charge transfer mechanisms differ significantly. Therefore, before examining the photocatalytic water-splitting performance of a van der Waals heterostructure, the operational mechanism of charge transfer should be established. One approach is to consider the direction of the interfacial built-in electric field and the differential charge density at the interface~\cite{Malefane2023}. These quantities alone are not enough to make a clear distinction between the two systems, because charge density results alone are not enough to give a conclusive answer.

More sophisticated strategies have been developed to monitor the real-time dynamics of the photogenerated charge carriers, such as non-adiabatic molecular dynamics (NAMD)~\cite{Qijing2019,Anqi2022}. However, NAMD is computationally expensive for large systems, and its accuracy can be limited due to neglecting spin-orbit coupling and using approximate exciton wave functions~\cite{Axelrod2022,Zhou2021}. Very recently, Liu et al.~\cite{Yaozhong2024} proposed a strategy based on the Gibbs free energy minimum barrier principle to overcome these limitations. In this strategy, the charge transfer mechanism is identified by comparing the maximum energy barriers of the oxygen evolution reaction (OER) and hydrogen evolution reaction (HER) between the type-II and Z-scheme configurations in the light-off state~\cite{Yaozhong2024}.  However, since charge transfer inherently involves the movement of electrons and photogenerated holes, it is more appropriate to evaluate this mechanism under light irradiation (light-on state) rather than in the dark (light-off state). To address this issue, Pingak et al.~\cite{Redi2025} extended the approach to the light-on state, where the Gibbs free energy differences of HER and OER intermediates are calculated on the different surfaces of the heterostructure for the two possible mechanisms under light irradiation. According to this method, the leading mechanism is the one that produces the smallest change in Gibbs free energy for the OER steps and keeps the hydrogen adsorption free energy ($\Delta G_{H^*}$) for the HER step as near to zero as possible, following the Sabatier principle. Here, we employ the same technique that offers a more reliable and physically meaningful criterion to identify the charge transfer process in heterostructures with staggered bands.

The water splitting process can fundamentally take place in both acidic and alkaline media. Most experimental investigations are performed in alkaline environments~\cite{Jakhar2021}, whereas the predominant focus of computational studies has been on acidic conditions, resulting in diminished attention to water oxidation in alkaline media~\cite{Congling2019,Jakhar2021}. This divergence in emphasis between theoretical and experimental research highlights the necessity for an exhaustive investigation of water splitting mechanisms under both acidic and alkaline environments.

We aim to achieve this goal on an exemplary GeC/ReS$_2$ heterobilayer. Germanium carbide (g-GeC) monolayers have received substantial interest for their potential uses in photocatalysis owing to their appropriate band gap, high dynamic and thermal stability, and excellent mechanical properties~\cite{Luo2019,Vu2020,Ji2018,Ferdous2023}. Previous studies have demonstrated that g-GeC can be used to form high-performance heterostructures with a variety of semiconductors, such as g-GeC/PtSe$_2$~\cite{Xu2023}, PtS$_2$/GeC~\cite{Ding2024}, GeC/SnSSe~\cite{Jiang2022}, MoSTe/g-GeC~\cite{Ma2024}, ZnO/GeC~\cite{Wang2019}, SiS/GeC~\cite{Ullah2023}, GeC/$\beta$AsP~\cite{Mao2023}, ZnTe/GeC~\cite{Cao2021} and GeC/AlN~\cite{Ferdous2023}.
Meanwhile, monolayer ReS$_2$ has also shown a wide light absorption from the visible to the ultraviolet region and good carrier mobility, making it a viable candidate for photocatalytic water splitting~\cite{Ho1998,Friemelt1992,Wang2023,Liu2016}. In fact, heterostructures based on ReS$_2$ have been extensively investigated. For instance, Z-scheme band alignment was demonstrated in AlN/ReS$_2$~\cite{Zhao2024}, and type-II configuration was confirmed in GaN/ReS$_2$~\cite{Wang2023}.

Motivated by the promising demonstration of g-GeC for heterostructure engineering, the well-established photocatalytic activity of ReS$_2$, and the absence of a systematic investigation of the GeC/ReS$_2$ heterostructure, we study its electronic structure, interfacial charge transfer, and photocatalytic activity using first-principles computations. Moreover, the Gibbs free energy variations of the HER and OER intermediates on the GeC and ReS$_2$ surfaces are computed under light irradiation in acidic and alkaline conditions. In the Z-scheme mechanism, OER and HER occur on the oxidation and reduction catalyst surfaces, respectively, whereas the reverse is the case for the type-II mechanism. The differences in their Gibbs free energies under light irradiation are helpful to determine the most favourable charge transfer pathway. We intend to gain a understanding of the charge-separation and redox-activity mechanisms in the GeC/ReS$_2$ heterostructure and to evaluate its potential for efficient water-splitting photocatalysis.

\section{Computational Methods}\label{sec:Computational Methods}
All computations were performed within the framework of the density functional theory (DFT)~\cite{Kohn1965,Hohenberg1964} as implemented in the Vienna ab initio simulation package (VASP)~\cite{Kresse1993,Kresse1996}. The interaction between core and valence electrons was handled by the projector augmented-wave (PAW) method~\cite{Blochl1994,Kresse1999}. The exchange-correlation functional was calculated by the generalized gradient approximation (GGA) of Perdew-Burke-Ernzerhof (PBE)~\cite{Perdew1996,Perdew1997} in the structural optimizations, while the hybrid functional of Heyd-Scuseria-Ernzerhof (HSE06)~\cite{Heyd2003} was used to obtain more accurate electronic and optical properties. We introduced a 15~\AA{} vacuum spacer in the perpendicular direction to avoid spurious interlayer interactions. The plane-wave kinetic energy cutoff was selected at 550~eV. The structural relaxation was considered converged when the energy variation was less than $10^{-5}$~eV and the residual atomic forces were less than $0.01$~eV/\AA{}. Long-range van der Waals interactions were taken into account by means of the DFT-D3~\cite{Grimme2010} correction.

For the Brillouin zone sampling, Monkhorst-Pack~\cite{Monkhorst1976} meshes of $10 \times 10 \times 1$ for the GeC monolayer, $5 \times 5 \times 1$ for the ReS$_2$ monolayer, and $5 \times 5 \times 1$ for the GeC/ReS$_2$ heterostructure were used. Further dynamic stability analysis was done for lattice vibrations in a $4 \times 4 \times 1$ supercell. Atomic displacements were generated with the PHONOPY code~\cite{TOGO2015} and the corresponding interatomic force constants were extracted and refined using the HIPHIVE package~\cite{Eriksson2019}. Furthermore, ab initio molecular dynamics (AIMD) simulations were performed in the canonical ensemble (NVT) using a Nosé-Hoover thermostat at 300 K. Finally, the data analysis and post-processing were performed utilizing the utility VASPKIT~\cite{VEI-WANG2021}.

\section{Results and Discussion}\label{sec:results} 

\subsection{Geometric Properties of GeC and ReS$_2$ Monolayers and Their Heterostructure Configuration}

\begin{figure}[ht!]
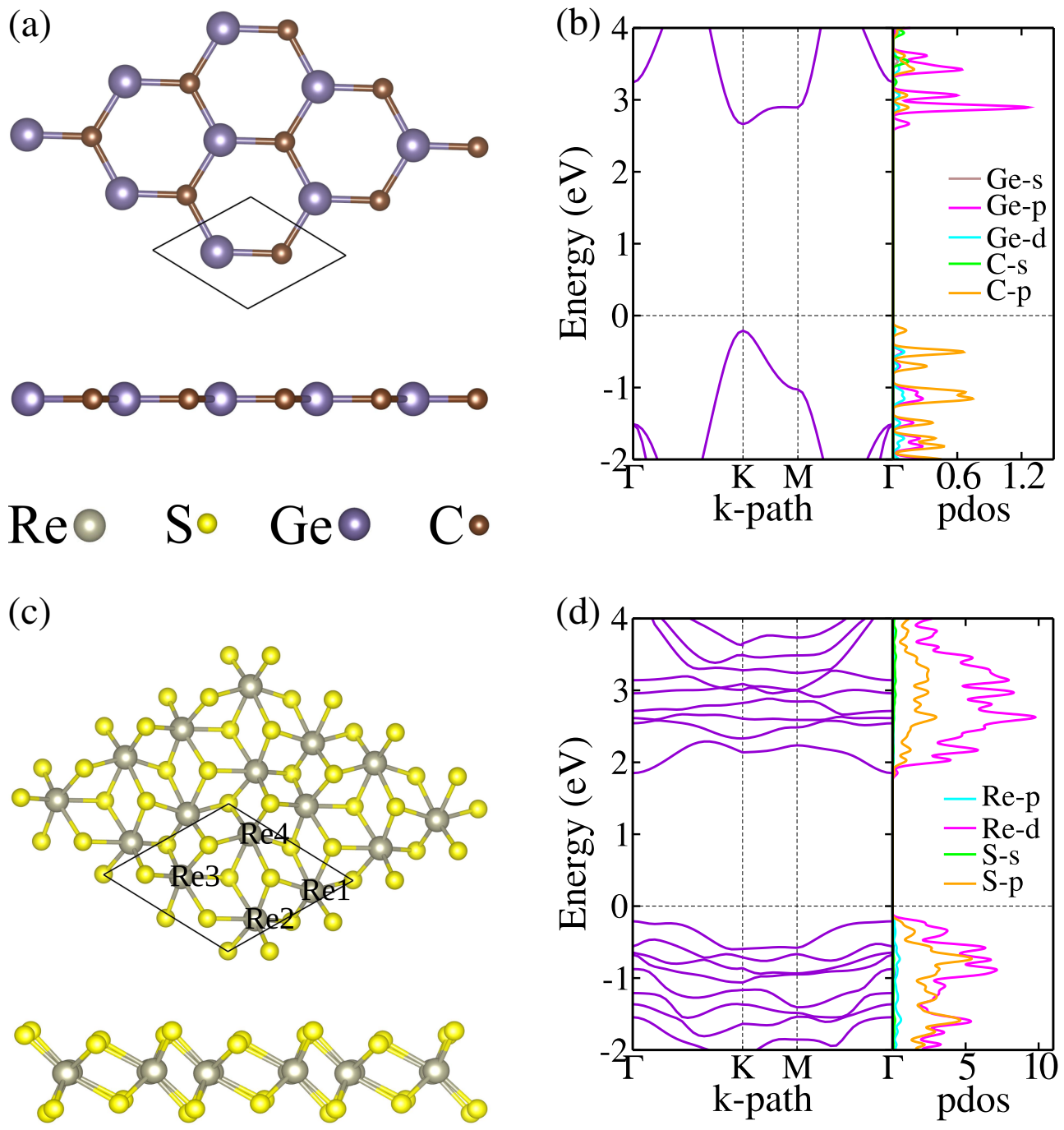

\centering
\begin{overpic}[scale=0.045]{./monolayers.pdf}
\end{overpic}
\caption{Structural and electrical characteristics of monolayers of ReS$_2$ and GeC (a) Top and side views of GeC monolayer. (b) Band structure and PDOS of GeC monolayer computed with HSE06. (c) Top and side view of ReS$_2$ monolayer. (d) Band structure and PDOS of ReS$_2$ monolayer with HSE06.}
\label{fig:monolayers_structures}
\end{figure}

Before the construction of the GeC/ReS$_2$ heterostructure, we investigated the single layers of GeC and ReS$_2$. The GeC monolayer is a flat honeycomb structure as illustrated in Fig.~\ref{fig:monolayers_structures}(a). The unit cell of the GeC monolayer contains one Ge atom and one C atom. The lattice constants and bond length of GeC monolayer are $a=b=3.23$~\AA{} and $d_{\mathrm{Ge-C}}=1.87$~\AA{}, respectively. The atomic structure of ReS$_2$ monolayer is given in Fig.~\ref{fig:monolayers_structures}(c). The ReS$_2$ monolayer is made of a layer of Re atoms sandwiched between two layers of S atoms. The layer contains four Re atoms and eight S atoms in the unit cell with lattice constants of $a=6.41$~\AA{} and $b=6.51$~\AA{} and Re-Re bond lengths $d_{\mathrm{Re1-Re2}}=2.82$~\AA{}, $d_{\mathrm{Re1-Re3}}=2.81$~\AA{}, $d_{\mathrm{Re2-Re3}}=2.72$~\AA{} and $d_{\mathrm{Re3-Re4}}=2.81$~\AA{}. The data are summarized in Table~\ref{geometry} and are in good agreement with prior studies~\cite{Yu2015, Yu2016}.

The band structure and partial density of states (PDOS) of GeC monolayer are shown in Fig.~\ref{fig:monolayers_structures}(b). The band structure analysis indicates that this single layer has a direct band gap of $2.87$~eV at the K point, and the PDOS of this monolayer shows that the valence band is dominated by the p-orbital of the C atom and the conduction band is dominated by the p-orbital of the Ge atom. The band structure and PDOS of the ReS$_2$ monolayer are presented in Fig.~\ref{fig:monolayers_structures}(d). As shown in Fig.~\ref{fig:monolayers_structures}(d), this monolayer has a direct band gap of $2.06$~eV at the $\Gamma$ point, and its PDOS demonstrates that the d-orbital of the Re atom has the biggest contribution in the valence and conduction bands. The electronic band gaps of GeC and ReS$_2$ are computed by HSE06 functional, and the resultant values are in good agreement with recent computational results~\cite{Islam2021,Wang2023}.

\begin{table*}
\caption{\label{geometry}
Structural parameters and energetics of monolayer GeC, monolayer ReS\textsubscript{2}, and GeC/ReS\textsubscript{2} heterostructure configurations: lattice constants (\textit{a}, \textit{b}; Å), bond lengths (\textit{d}; Å), equilibrium energy (\textit{E}\textsubscript{0}; eV), binding energy (\textit{E}\textsubscript{b}; eV), and interlayer distance (\textit{h}; Å).}
\setlength{\tabcolsep}{8pt}
\begin{tabular}{c|cccccccccc}
\hline
\hline
System & a  & b & $d_\text{Ge-C}$ & $d_\text{Re1-Re2}$ & $d_\text{Re1-Re3}$  & $d_\text{Re2-Re3}$ & $d_\text{Re3-Re4}$ & $E_0$ & $E_\text{b}$ & h \\

\hline
GeC & 3.23 & 3.23 & 1.87 & - & - & - & - & -49.034 & - & - \\

ReS$_2$ & 6.41 & 6.51 & - & 2.82 & 2.81 & 2.72 & 2.81 & -92.890 & - & - \\

GeC/ReS$_2$ (I) & 6.41 & 6.46 &  1.86 & 2.82 & 2.78 & 2.70 & 2.82 & -142.766 & -0.842 & 3.04 \\

GeC/ReS$_2$ (II) & 6.40 & 6.46 & 1.86 & 2.82 & 2.78 & 2.70 & 2.82 & -142.809 & -0.885 & 3.01 \\

GeC/ReS$_2$ (III) & 6.40 & 6.46 & 1.86 & 2.82 & 2.78 & 2.70 & 2.82 & -142.793 & -0.869 & 3.02 \\

GeC/ReS$_2$ (IV) & 6.40 & 6.46 & 1.86 & 2.82 & 2.78 & 2.70 & 2.82 & -142.827 & -0.903 & 2.96 \\

\hline
\hline
\end{tabular}
\label{tab:results}
\end{table*}

\begin{figure*}[ht!]
\centering
\begin{overpic}[scale=0.75]{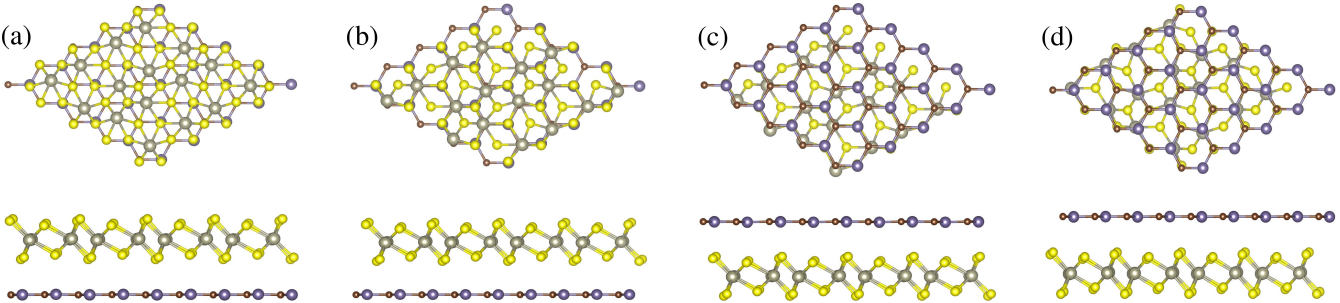}
\end{overpic}
\caption{Four distinct stacking patterns of the GeC/ReS$_2$ van der Waals heterostructure: (a) Configuration I, (b) Configuration II, (c) Configuration III, (d) Configuration IV.}
\label{fig:structures}
\end{figure*}

The first step in making a heterostructure is to calculate the lattice mismatch between the two monolayers. The lattice mismatch (LMM) between the two monolayers was computed as~\cite{Bafekry2021}:
\begin{equation}
    \text{LMM}=\left| \frac{a_{\text{GeC}} - a_{\text{ReS$_2$}}}{a_{\text{GeC}}} \right| \times 100\% ,
\end{equation}
where $a_{\text{GeC}}$ and $a_{\text{ReS$_2$}}$ are the optimized lattice constants of the GeC and ReS$_2$ monolayers, respectively. A lattice mismatch of less than 5\% confirms the possibility of experimental synthesis of this structure~\cite{Hu2019}. The lattice constant of the ReS$_2$ layer is $\sim$2 times larger than the lattice constant of the GeC layer. Thus, a $2 \times 2$ supercell of the GeC layer is joined with a unit cell of ReS$_2$ to construct the GeC/ReS$_2$ heterostructure, which has a lattice mismatch of 0.7\% in the a and b directions, and is ideal for heterostructure fabrication.

We have explored four possible configurations of the two monolayer arrangements to construct the GeC/ReS$_2$ heterostructure, and their geometric structures are illustrated in Fig.~\ref{fig:structures}. The interlayer distances of the combinations I, II, III, and IV were 3.04~\AA{}, 3.01~\AA{}, 3.02~\AA{}, and 2.96~\AA{}, respectively. We have calculated the binding energy of the GeC/ReS$_2$ heterostructure using~\cite{Cheng2022,Khawla2024} as:
\begin{equation}
    E_b=E_{\text{GeC/ReS$_2$}} - E_{\text{GeC}} - E_{\text{ReS$_2$}},
\end{equation}
Here, $E_{\text{GeC}}$ and $E_{\text{ReS$_2$}}$ are the energies of the monolayers of GeC and ReS$_2$, and $E_{\text{GeC/ReS$_2$}}$ is the total energy of the GeC/ReS$_2$ heterostructure. The binding energies of these configurations were $-0.842$~eV, $-0.885$~eV, $-0.869$~eV and $-0.903$~eV respectively. The results are presented in Table~\ref{tab:results}. The negative binding energy values indicate that the heterostructure is stable and the more negative the value of the binding energy, the more thermodynamically stable the system is. Hence, the lowest binding energy and shortest interlayer distance were found to be the most stable configuration, which is configuration IV. This configuration was therefore chosen for further investigation and various computations were performed on it.

\subsection{Stability Analysis of the GeC/ReS$_2$ Heterostructure}
\begin{figure}[h!]
\centering
\includegraphics*[scale=0.22]{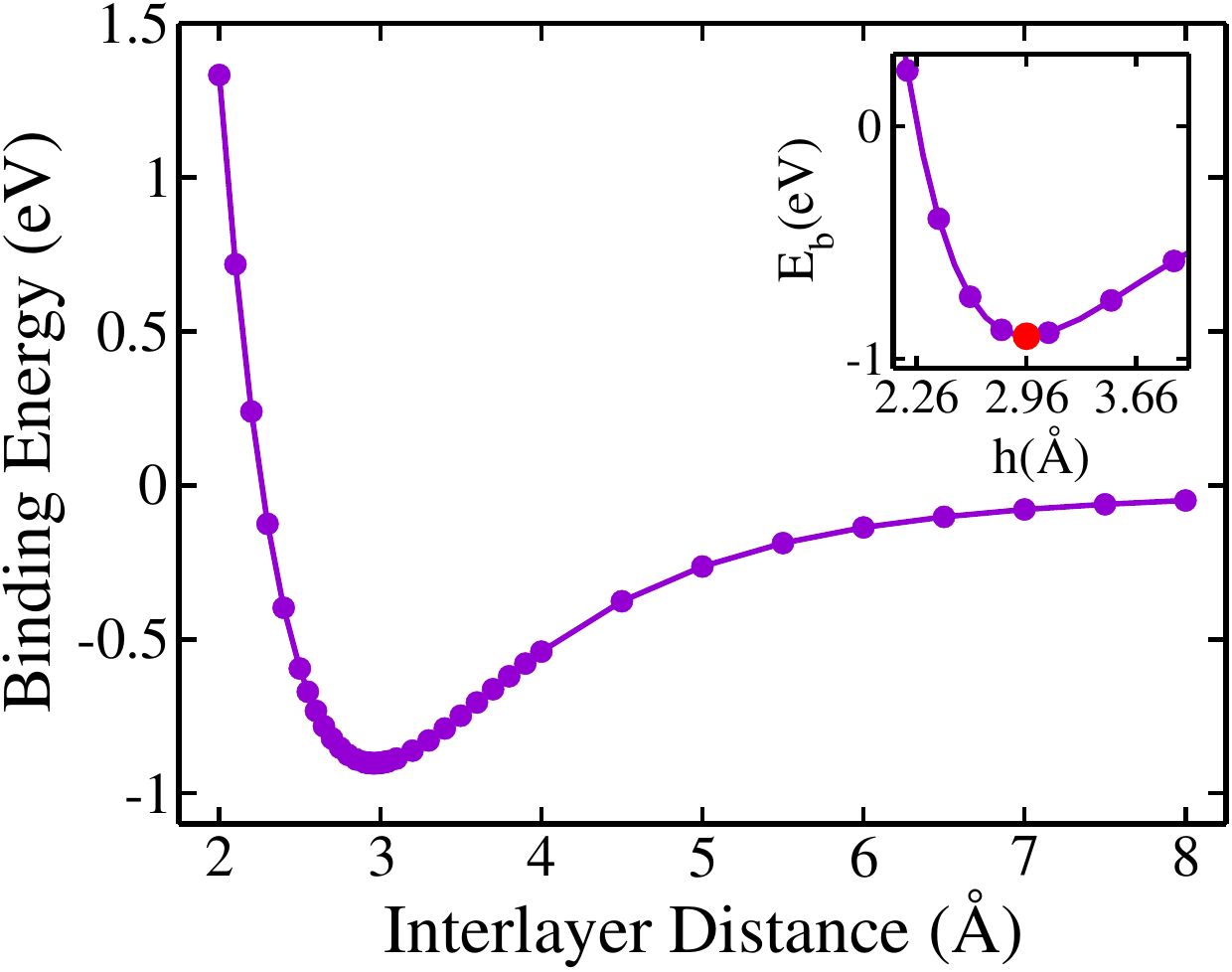}
\caption{Binding energy as a function of interlayer distance for GeC/ReS$_2$ heterostructure. The inset shows the optimal atom arrangement at the equilibrium point.}
\label{fig:binding_energy}
\end{figure}

To study the thermodynamic stability of the GeC/ReS$_2$ heterostructure, the variations of binding energy with interlayer distance were investigated by using the DFT-D3 van der Waals functional. From the binding energy curve of the most stable configuration (see Fig.~\ref{fig:binding_energy}), the ideal interlayer distance of the GeC/ReS$_2$ heterostructure is 2.96~\AA{}, where the binding energy is minimized and the thermodynamic stability is the greatest.

\begin{figure}[ht!]
\centering
\begin{overpic}[scale=0.173]{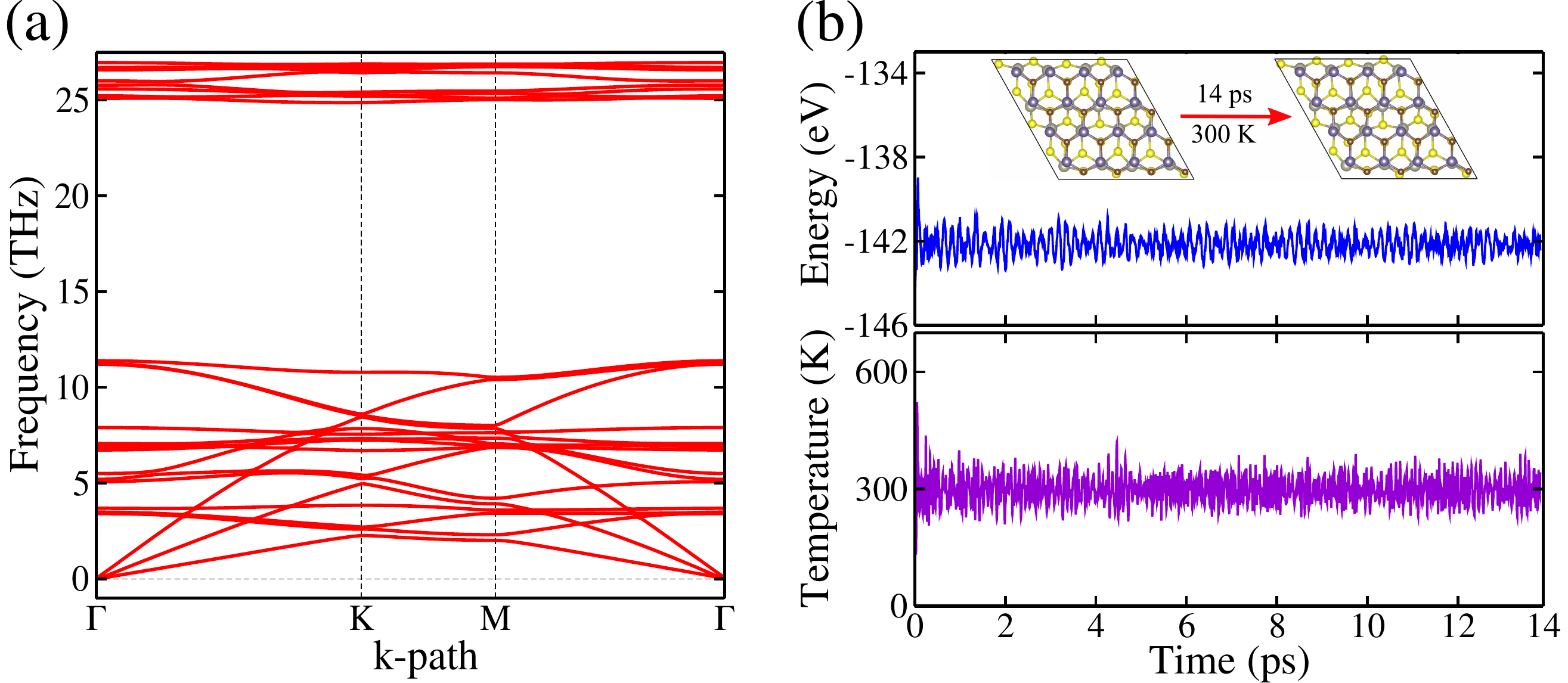}
\end{overpic}
\caption{(a) Phonon dispersion and (b) AIMD simulation at 300 K for the GeC/ReS$_2$ heterostructure.}
\label{fig:stability}
\end{figure}

 Fig.~\ref{fig:stability}(a) displays the phonon dispersion curve with no negative frequencies, suggesting its dynamic stability. 
The phonon bands show three acoustic modes, a longitudinal (LA) and in-plane transverse (TA), and an out-of-plane mode (ZA) with linear dispersion ($\omega \propto q$) for $q \rightarrow 0$ and no out-of-plane flexure mode (ZA) with quadratic dispersion in the long-wavelength limit, which is a common feature of 2D systems, such as graphene and arsenene~\cite{Sarsari2016}. Flexural modes are very important in contributing to the heat conductivity both as carriers and scatterers. We also notice a gap of roughly 22 THz between the low-frequency and high-frequency optical branches, which restricts the three-phonon scattering channels. The heavy-atom modes are blocked from scattering with phonons by the light-atom modes, and this can increase the lattice thermal conductivity.

Ab initio molecular dynamics (AIMD) simulations were carried out at a temperature of 300~K for 14 ps. As shown in Fig.~\ref{fig:stability}(b), no structural damage was observed throughout these 14~ps. The results show that the GeC/ReS$_2$ heterostructure has good thermal stability.

\subsection{Electronic Properties of the GeC/ReS$_2$ Heterostructure}

\begin{figure}[ht!]
\centering
\begin{overpic}[scale=0.265]{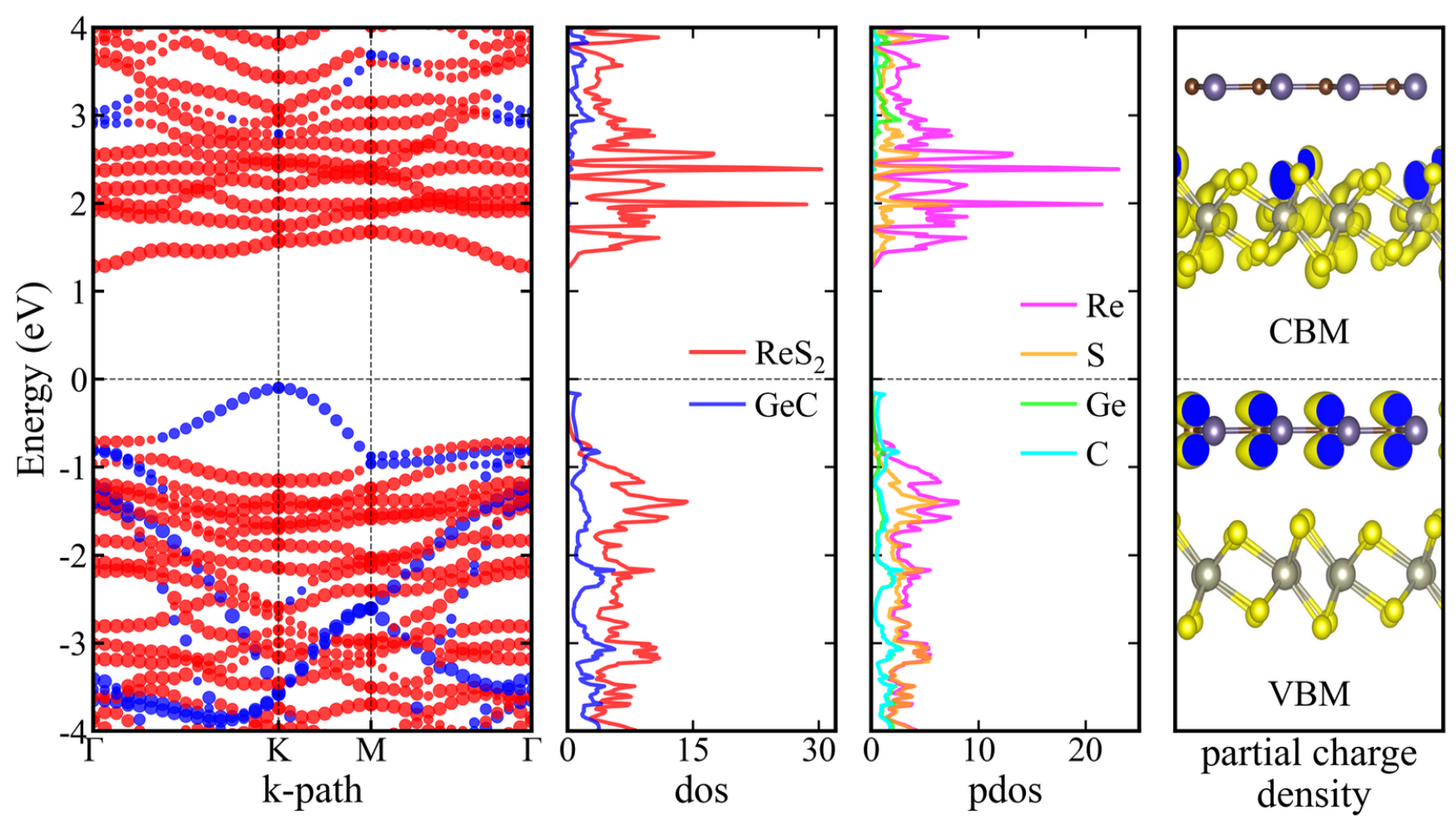}
\end{overpic}
\caption{Electronic features of the GeC/ReS$_2$ heterostructure: (a) Band structure, (b) density of states (DOS), (c) projected DOS and (d) band-decomposed charge density for the CBM and VBM.}
\label{fig:pband_pdos-HS}
\end{figure}

After validating the stability of the GeC/ReS$_2$ heterostructure, we now explore its water splitting performance. 
The electronic characteristics were analyzed to gain insight into the mechanism of water splitting. The calculated band structure (Fig.~\ref{fig:pband_pdos-HS}(a)) exhibits an indirect band gap of 1.38 eV in the heterostructure. The valence band maximum (VBM) is placed at the K point, while the conduction band minimum (CBM) is located at the $\Gamma$ point. The VBM consists of the GeC layer with contributions from the ReS$_2$ layer at the bottom of the valence band. A similar situation occurs for the CBM where the ReS$_2$ layer contributes, and the GeC layer contribution can also be seen in the higher regions of the conduction band.

The DOS and PDOS plots (Figs.~\ref{fig:pband_pdos-HS}(b) and (c)) show that the states near VBM are mostly from C atoms in the GeC layer, while the states near CBM are mainly from Re and S atoms in the ReS$_2$ layer. The band-decomposed charge density (Fig.~\ref{fig:pband_pdos-HS}(d)) also shows that the total charge density of VBM is associated with the GeC layer, while the total charge density of CBM is associated with the ReS$_2$ layer. The electronic characteristics of the GeC/ReS$_2$ heterostructure (Figs.~\ref{fig:pband_pdos-HS}(a)-(d)) clearly demonstrate the spatial separation of the valence- and conduction-band edges of the two monolayers, showing a staggered band alignment. Such a staggered band alignment is a common feature of both type-II and Z-scheme heterostructures.

\subsection{Interfacial Charge Transfer and Built-in Electric Field}

As mentioned earlier, the analysis of the interfacial charge transfer and the built-in electric field provides the basic first insight into the charge transfer mechanism of a heterostructure. 
The fundamental theoretical distinction between the type-II and direct Z-scheme heterostructures resides in the charge transfer pathway, dictated by the direction of the produced internal electric field \cite{Malefane2023}. In a  type-II heterostucture, the Fermi level of the oxidation semiconductor is greater than that of the reduction semiconductor, and the direction of the induced internal electric field is from the oxidation semiconductor to the reduction semiconductor. Therefore, the bending of the valence and conduction bands at the interface of the type-II heterostructure enables the movement of electrons from the higher conduction band to the lower conduction band and holes from the lower valence band to the higher valence band, leading to a uniform Fermi level. In addition, a  type-II heterostructure suppresses the recombination between electrons in the conduction band of the oxidation semiconductor and holes in the valence band of the reduction semiconductor. Instead, these electrons and holes participate in the water-splitting process.

Conversely, in a direct Z-scheme heterostructure, the Fermi level of the reduction semiconductor is greater than that of the oxidation semiconductor, and the direction of the induced internal electric field is from the reduction semiconductor to the oxidation semiconductor. The bending of the valence and conduction bands at the interface of the Z-scheme heterostructure causes direct charge transfer from the conduction band of the oxidation layer to the valence band of the reduction layer, leading to a uniform Fermi level. Thus, the selective recombination in the Z-scheme heterostructure occurs between the conduction band electrons of the oxidation semiconductor and the valence band holes of the reduction semiconductor. This selective recombination prohibits the transport of electrons from the higher conduction band to the lower conduction band and the transfer of holes from the lower valence band to the higher valence band. Instead, in a Z-scheme heterostructure, electrons in the conduction band of the reduction semiconductor and holes in the valence band of the oxidation semiconductor participate in the water splitting process.

\begin{figure}[ht!]
\centering
\begin{overpic}[scale=0.19]{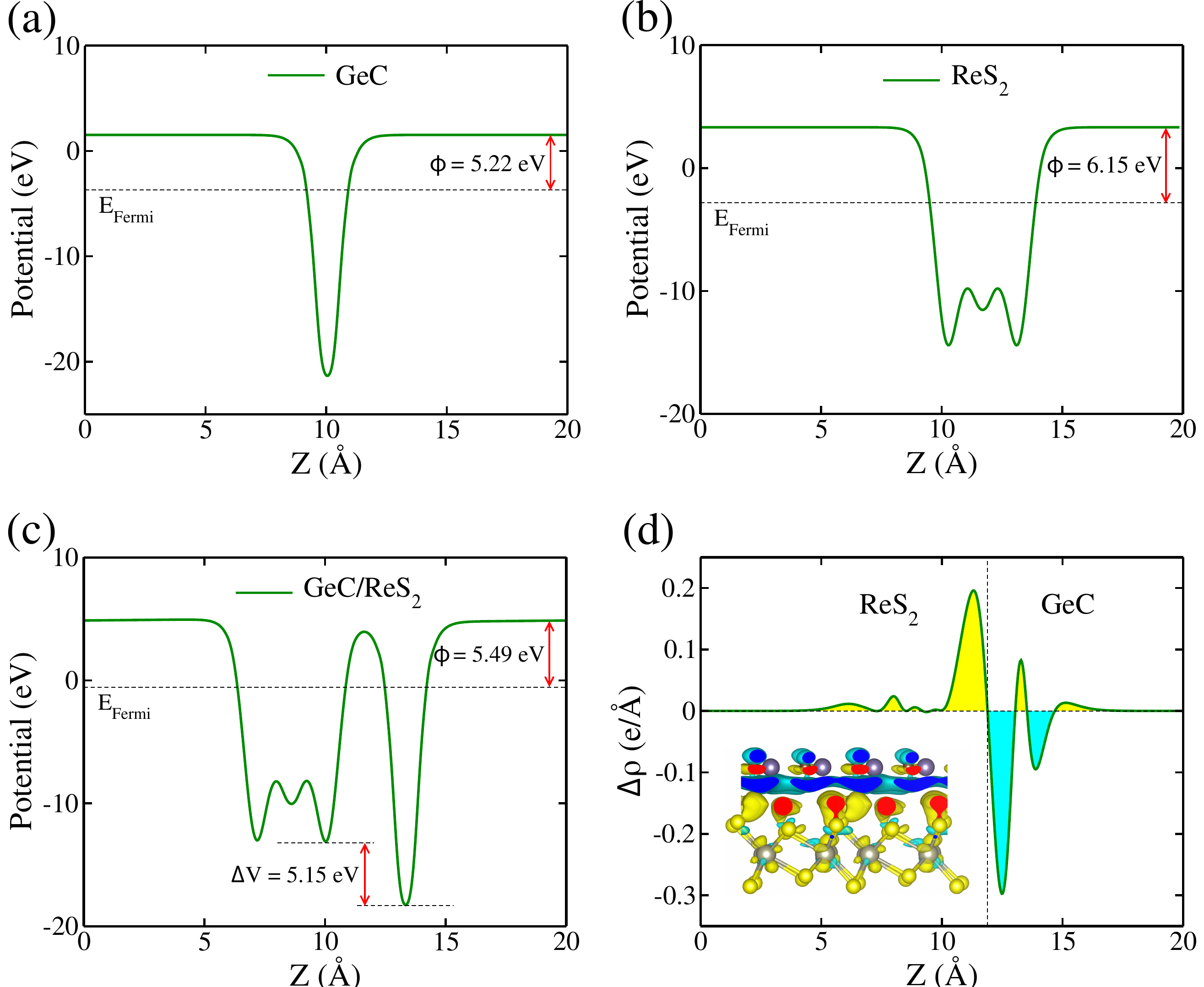}
\end{overpic}
\caption{Distribution of electrostatic potential along the z-direction of (a) monolayer GeC, (b) monolayer ReS$_2$ and (c) GeC/ReS$_2$ heterostructure. (d) Plane averaged charge density difference for the GeC/ReS$_2$ heterojunction.}
\label{fig:potentials}
\end{figure}

To better understand the origin and mechanism of charge transfer at the interface of GeC/ReS$_2$ heterostructure, the average electrostatic potential of the GeC monolayer, ReS$_2$ monolayer and GeC/ReS$_2$ heterostructure and the charge density difference of GeC/ReS$_2$ heterostructure (Fig.~\ref{fig:potentials}) were calculated.
The work function ($\phi$) is an important parameter for the band alignment, which is given by the following equation~\cite{Liu2015,Jianjun2015}: 
\begin{equation}
    \phi=E_{\text{vacuum}} - E_{\text{Fermi}} ,
\end{equation}
Here $E_{\text{Fermi}}$ is the Fermi level and $E_{\text{vacuum}}$ is the vacuum level. 
The work function ($\phi$) is the minimum energy to move an electron from the Fermi surface to the vacuum, and can forecast the charge transfer direction across layers during the formation of the GeC/ReS$_2$ heterostructure~\cite{Liu2015,Jianjun2015}.

As indicated in the average electrostatic potential diagrams (Figs.~\ref{fig:potentials}(a)-(c)), the work functions of the single layers of GeC and ReS$_2$ are 5.22 and 6.15 eV, respectively, whereas the work function of the GeC/ReS$_2$ heterostructure is 5.49 eV. Furthermore, the average electrostatic potential diagram of GeC/ReS$_2$ heterostructure exhibits a notable potential difference of $\Delta$V=5.15 eV at the interface of GeC and ReS$_2$. This potential difference causes charge transfer at the interface and drives spontaneous electron transfer from the GeC layer (lower work function) to the ReS$_2$ layer (higher work function). Such interfacial charge transfer results in a positive net charge of the GeC layer and a negative net charge of the ReS$_2$ layer. A substantial potential difference suggests the formation of a strong built-in electric field in the interface from GeC to ReS$_2$.

To characterize the charge transfer and distribution at the interface of the GeC/ReS$_2$ heterostructure, the charge density difference (CDD) is calculated as~\cite{Tang2014,Preuss2005}:  
\begin{equation}
    \Delta \rho=\rho_{\text{GeC/ReS$_2$}} - \rho_{\text{GeC}} - \rho_{\text{ReS$_2$}} ,
\end{equation}
where $\Delta \rho$ is the charge density difference of the GeC/ReS$_2$ heterostructure, $\rho_{\text{GeC/ReS$_2$}}$ is the total charge density of the GeC/ReS$_2$ heterostructure, $\rho_{\text{GeC}}$ and $\rho_{\text{ReS$_2$}}$ are the charge densities of the GeC and ReS$_2$ monolayers, respectively. The CDD offers information on the direction and amplitude of charge transfer between the layers and indicates the nature of interactions at the interface of the layers.

As seen in the charge density difference plot (Fig.~\ref{fig:potentials}(d)), the ReS$_2$ monolayer shows the positive values of $\Delta\rho$, whereas the GeC monolayer largely shows the negative values of $\Delta\rho$. Thus, the charge buildup is mostly at the ReS$_2$ layer, whereas the charge depletion is principally at the GeC layer at the interface of the GeC/ReS$_2$ heterostructure. This suggests charge transfer from GeC to ReS$_2$ and validates the presence of a built-in electric field from GeC to ReS$_2$. Thus, the charge density difference in Fig.~\ref{fig:potentials}(d) is in good agreement with the average electrostatic potential results in Fig.~\ref{fig:potentials}(c) and provides convincing evidence for the direction of charge transfer and the built-in electric field in the GeC/ReS$_2$ heterostructure. These results are in agreement with the Z-scheme charge transfer mechanism, which will be further confirmed by the upcoming Gibbs free energy study.

\subsection{Band Alignment and Charge Transfer Pathways in the GeC/ReS$_2$ Heterostructure}

Having established the direction of the built-in electric field, we now discuss the band alignment of the GeC/ReS$_2$ heterostructure. This allows us to assess the thermodynamic feasibility for overall water splitting, and demonstrates the charge transfer channels obtained.
For successful overall water splitting, the conduction band minimum and valence band maximum of a semiconductor should be well aligned in relation to the water redox potentials. Specifically, the VBM should be at or below the oxidation potential to generate enough holes for the oxidation process, and the CBM should be at or above the reduction potential to drive the reduction process. The absolute positions of VBM and CBM relative to the vacuum level are calculated using~\cite{QIANG2024,Fei2022}:
\begin{equation}
E_{\text{VBM}}=E_{\text{VBM}}^{\text{HSE06}} - E_{\text{vacuum}} ,
\end{equation}
\begin{equation}
E_{\text{CBM}}=E_{\text{CBM}}^{\text{HSE06}} - E_{\text{vacuum}} ,
\end{equation}
where $E_{\text{VBM}}^{\text{HSE06}}$ and $E_{\text{CBM}}^{\text{HSE06}}$ are the band edge energies from the HSE06 functional. To evaluate the effectiveness for photocatalytic water splitting, the positions of these bands relative to the vacuum level need to be determined.
The standard reduction potential for the HER and the standard oxidation potential for the OER can be taken from~\cite{Liu2011,Zhuang2013} as: 
\begin{equation}
    E_{\text{red}}=-4.44 \,\text{eV} + 0.059 \times \text{pH} ,
\end{equation}
\begin{equation}
    E_{\text{ox}}=-5.67 \,\text{eV} + 0.059 \times \text{pH} ,
\end{equation}
These redox potentials set the thermodynamic boundaries for the water splitting half-reactions. Since these potentials are very pH dependent, the values at pH=0, 7 and 14 are calculated to be -5.67, -5.257 and -4.844 eV for oxidation and -4.44, -4.027 and -3.614 eV for reduction, respectively.

\begin{figure*}[ht!]
\centering
\begin{overpic}[scale=0.5]{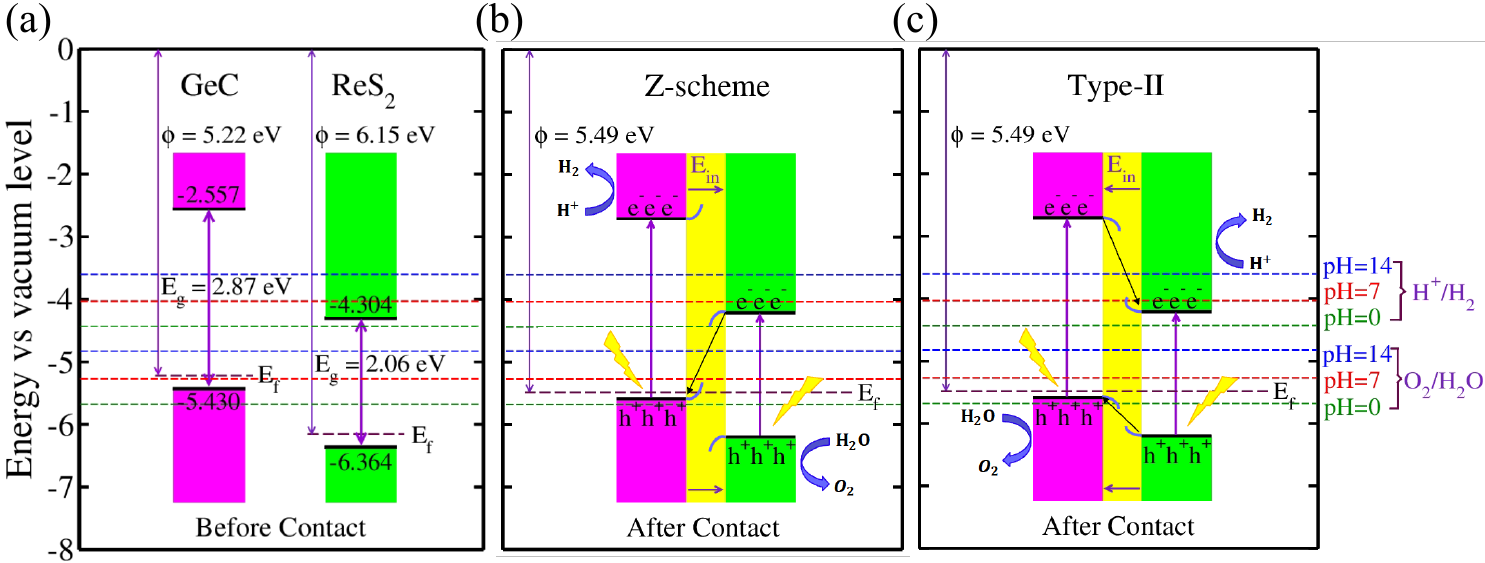}
\end{overpic}
\caption{Band alignment: (a) Monolayers (GeC and ReS$_2$), (b) Z-scheme pathway, and (c) Type-II pathway of GeC/ReS$_2$ heterostructure.}
\label{fig:bandalignment}
\end{figure*}

Fig.~\ref{fig:bandalignment} shows the band-edge energies of GeC and ReS$_2$ monolayers, the Z-scheme mechanism in the GeC/ReS$_2$ heterostructure, and the type-II mechanism in the GeC/ReS$_2$ heterostructure.
These are compared to water redox potentials at pH=0 (green lines), pH=7 (red lines) and pH=14 (blue lines). As shown in Fig.~\ref{fig:bandalignment}(a), the VBM and CBM energies of the GeC monolayer are -5.430 eV and -2.557 eV, respectively. This configuration enables the GeC monolayer to catalyze the HER over the whole pH range (0–14), whereas the OER activity is restricted to pH range of 4-14 and is thermodynamically banned at pH 0-3. The VBM and CBM energies of the ReS$_2$ monolayer are -6.364 eV and -4.304 eV, respectively. Consequently, the ReS$_2$ monolayer can catalyze the OER in the whole pH range (0–14), while the HER is only active in the pH range of 0–2 and inactive in the pH range of 3–14. Note that the computed band edge positions of both monolayers are in good agreement with prior theoretical data~\cite{TIAN2024,Zhao2024}.

After constructing the GeC/ReS$_2$ heterostructure, the band edges of the two layers are slightly shifted due to the weak van der Waals interactions. The VBM and CBM of the GeC layer shift to -5.589 eV and -2.705 eV, respectively, whereas the VBM and CBM of the ReS$_2$ layer shift to -6.202 eV and -4.210 eV, respectively. Under the light irradiation, the electrons are excited from the valence to the conduction bands of both layers, and therefore, the electron-hole pairs are generated.

The Z-scheme mechanism is shown in Fig.~\ref{fig:bandalignment}(b). The electrons transfer from GeC to ReS$_2$ to match the Fermi levels, which results in a built-in electric field at the interface. The potential barrier produces band bending, shifting the GeC bands upward and the ReS$_2$ bands downward. This band bending preferentially promotes recombination of electrons in the ReS$_2$ conduction band with holes in the GeC valence band. At the same time, it acts as a barrier for the recombination of GeC conduction electrons with ReS$_2$ valence holes. This leads to electron accumulation in the CBM of GeC and hole accumulation in the VBM of ReS$_2$ correspondingly. This spatial separation considerably minimizes the recombination of charges and promotes the photocatalytic performance. Finally, the HER is driven by the accumulating electrons in GeC, and the holes in ReS$_2$ drive the OER. This Z-scheme charge transfer mechanism can achieve total water splitting across a wide pH range (0–14).
By contrast, the type-II mechanism (Fig.~\ref{fig:bandalignment}(c)) possesses an opposite-direction internal electric field. This field pushes holes from the VBM of ReS$_2$ to the VBM of GeC and electrons from the CBM of GeC to the CBM of ReS$_2$. In this situation, the band bending inhibits the recombination of conduction electrons in ReS$_2$ with valence holes in GeC. We find the GeC valence holes to be involved in the OER while the ReS$_2$ conduction electrons are involved in the HER. However, the poor alignment of these charge carriers with the water redox potentials greatly reduces the overall redox driving force. The OER is thermodynamically not favourable at pH 0 and 1 but viable in the pH range of 2-14 in this type-II arrangement. Conversely, the HER is only feasible in the pH range of 0–3 and cannot occur in higher pH values (4–14). 

The Z-scheme mechanism achieves total water splitting in a wide pH range (0–14) due to the built-in electric field and the advantageous band alignment, which distinguishes it from the type-II mechanism. 
The thermodynamic viability of these two charge transfer mechanisms is further investigated by calculating and analyzing the Gibbs free energy changes of the HER and OER intermediates in the subsequent section.

\subsection{Photocatalytic Properties of the GeC/ReS$_2$ Heterostructure}
Gibbs free energy ($\Delta G$) is one of the most important thermodynamic properties to evaluate the spontaneity of the photocatalytic process. Thus, the change in the Gibbs free energy of each intermediate state in HER and OER for the type-II and Z-scheme charge transfer mechanisms is explored to gain better insight into the photocatalytic process in the GeC/ReS$_2$ heterostructure to assess their thermodynamic performance by the following formula~\cite{Tuan2024,Anjum2024,Li2021}:
\begin{equation}
    \Delta G=\Delta E + \Delta ZPE - T\Delta S + G_U - G_\text{pH}
\end{equation}
where $\Delta E$ is the change in adsorption energy of each intermediate step calculated using DFT calculations and T=298.15 K. $\Delta {ZPE}$ is the difference in zero-point energy. $\Delta S$ is the change in entropy between the intermediate species adsorbed on the surface of the photocatalyst and in their gaseous form. Both quantities are computed from vibrational frequency calculations.

The $G_\text{pH}$ term considers the concentration of protons in the solution. It is calculated as $G_\text{pH}=0.059\times\text{pH}$~\cite{Anjum2024} and represents free energy dependence on the pH of the medium. The term $G_{U}$ is the influence of the photoinduced potential and is given by $G_U=-neU_{SHE}$~\cite{Anjum2024}, where n is the number of electron-hole pairs exchanged in the process. U is the photogenerated charge carrier potential (electron or hole) relative to the standard hydrogen electrode (SHE). The effective photogenerated electron potential (U$_e$) is the difference of the CBM and the hydrogen reduction potential, and the effective photogenerated hole potential (U$_h$) is the difference of the hydrogen reduction potential and the VBM.  

For spontaneous water splitting, the photogenerated carriers should possess adequate driving power for hydrogen evolution reaction and oxygen evolution reaction. When U$_e$ and U$_h$ are larger than $\Delta G_{H^*}$ and $\Delta G_{max}$ (along the OER pathway) respectively, HER and OER can spontaneously occur in the light-on state. The pH correction of these potentials at 298.15 K is represented by the following two relations~\cite{Tuan2024}:
\begin{equation}
    U_e=U_e(\text{pH=0}) - 0.059 \times \text{pH}
\end{equation}
\begin{equation}
    U_h=U_h(\text{pH=0}) + 0.059 \times \text{pH}
\end{equation}

The hydrogen evolution reaction is a half-reaction useful for the production of the H$_2$ molecule, which is based on the mechanism of the Volmer–Heyrovsky or Volmer–Tafel reactions\cite{Song2021}. In an acidic medium, these reactions proceed in the following form~\cite{Liang2021,Norskov2004}: 
\begin{equation}
    * + \mathrm{H^+} + e^- \rightarrow \mathrm{H^*}
\end{equation}
\begin{equation}
    \mathrm{H^*} + \mathrm{H^+} + e^- \rightarrow \mathrm{H_2} + *
\end{equation}
\begin{equation}
    2\mathrm{H^*} \rightarrow \mathrm{H_2} + *
\end{equation}

In alkaline media, the water molecule acts as the proton source instead of a proton and the steps are expressed as follows~\cite{Liang2021}:
\begin{equation}
    \mathrm{H_2O} + e^- \rightarrow \mathrm{H^*} + \mathrm{OH^-}
\end{equation}
\begin{equation}
    \mathrm{H^*} + \mathrm{H_2O} + e^- \rightarrow \mathrm{*} + \mathrm{H_2} + \mathrm{OH^-}
\end{equation}
\begin{equation}
    2\mathrm{H^*} \rightarrow \mathrm{H_2} + 2*
\end{equation}

Following these steps, the Gibbs free energy change of the hydrogen adsorption in both acidic and alkaline environments is determined by the following formula~\cite{Liang2021,Norskov2004}:
\begin{equation}
    \begin{aligned}
    \Delta &G_{H^*}=[E_{H^*} -\frac{1}{2}E_{H_2} - E_*] + [ZPE_{H^*} - \frac{1}{2}ZPE_{H_2}]\\ &\quad - [TS_{H^*} - \frac{1}{2}TS_{H_2}] + G_U - G_\text{pH}
\end{aligned}
\end{equation}
$\Delta G_{H^*}$ is frequently utilized as a crucial indicator to forecast the photocatalytic activity for HER. This value should be zero or near zero for a perfect photocatalyst, as a positive $\Delta G_{H^*}$ denotes that it is not favourable to adsorb hydrogen, and a large negative $\Delta G_{H^*}$ means that hydrogen does not desorb easily from the photocatalyst surface. Thus, the value of $\Delta G_{H^*}$ is one of the criteria to measure the attractiveness of the HER route in both the type-II and Z-scheme processes.

The water splitting process is composed of two halves, with the other part being the oxygen evolution reaction, in which the water molecule is oxidized and molecular oxygen is released. The entire oxidation reaction of water under acidic conditions (pressure 1 bar and temperature of 298.15 K) is given by~\cite{Liang2021,Norskov2004,ROSSMEISL2007}
\begin{equation}
    2\mathrm{H_2O(l) \rightarrow O_2 + 4\mathrm{H^+} + 4\mathrm{e^-}},\Delta G_0=4.92~\text{eV}
\end{equation}
This reaction in an acidic medium consists of four sequential steps~\cite{Liang2021,Norskov2004,ROSSMEISL2007}:
\begin{equation}
    2\mathrm{H_2O(l) + * \rightarrow \mathrm{OH^*} + \mathrm{H^+} + \mathrm{e^-}}
\end{equation}
\begin{equation}
    \mathrm{OH^*} \rightarrow \mathrm{O^*} + \mathrm{H^+} + \mathrm{e^-} 
\end{equation}
\begin{equation}
    \mathrm{O^*} + \mathrm{H_2O(l)} \rightarrow \mathrm{OOH^*} + \mathrm{H^+} + e^-
\end{equation}
\begin{equation}
    \mathrm{OOH^*} \rightarrow * + \mathrm{O_2(g)} + \mathrm{H^+} + e^-
\end{equation}

In an alkaline medium, the overall oxidation reaction of water and these four steps are written in the presence of hydroxide ions as follows~\cite{Liang2021}: 
\begin{equation}
    4\mathrm{OH^-} \rightarrow \mathrm{O_2(g)} + 2\mathrm{H_2O(l)} + 4e^- 
\end{equation}
\begin{equation}
    * + \mathrm{OH^-} \rightarrow \mathrm{OH^*} + e^-
\end{equation}
\begin{equation}
    \mathrm{OH^*} + \mathrm{OH^-} \rightarrow \mathrm{O^*} + \mathrm{H_2O(l)} + e^-
\end{equation}
\begin{equation}
    \mathrm{O^*} + \mathrm{OH^-} \rightarrow \mathrm{OOH^*} + e^- 
\end{equation}
\begin{equation}
    \mathrm{OOH^*} + \mathrm{OH^-} \rightarrow \mathrm{O_2(g)} + \mathrm{H_2O(l)} + e^-
\end{equation}
To calculate the Gibbs free energy change of each step of the OER in both acidic and alkaline media, the following relationships are used~\cite{Liang2021}: 
\begin{equation}
    \begin{aligned}
    \Delta &G_1=[E_{\mathrm{OH^*}} + \frac{1}{2}E_{H_2} - E_{H_2O} - E_*]\\ &\quad + [ZPE_{\mathrm{OH^*}} + \frac{1}{2}ZPE_{H_2} - ZPE_{H_2O}]\\ &\quad - [TS_{\mathrm{OH^*}} + \frac{1}{2}TS_{H_2} -TS_{H_2O}] \\ &\quad + G_U - G_\text{pH}
    \end{aligned}
\end{equation}
\begin{equation}
    \begin{aligned}
    \Delta &G_2=[E_{\mathrm{O^*}} + \frac{1}{2}E_{H_2} - E_{\mathrm{OH^*}}]\\ &\quad + [ZPE_{\mathrm{O^*}} + \frac{1}{2}ZPE_{H_2} - ZPE_{\mathrm{OH^*}}]\\ &\quad - [TS_{\mathrm{O^*}} + \frac{1}{2}TS_{H_2} - TS_{\mathrm{OH^*}}] \\ &\quad + G_U - G_\text{pH}
\end{aligned}
\end{equation}
\begin{equation}
    \begin{aligned}
    \Delta &G_3=[E_{\mathrm{OOH^*}} + \frac{1}{2}E_{H_2} - E_{\mathrm{O^*}} - E_{H_2O}]\\ &\quad + [ZPE_{\mathrm{OOH^*}} + \frac{1}{2}ZPE_{H_2} - ZPE_{\mathrm{O^*}} - ZPE_{H_2O}]\\ &\quad - [TS_{\mathrm{OOH^*}} + \frac{1}{2}TS_{H_2} - TS_{\mathrm{O^*}} -TS_{H_2O}]\\ &\quad + G_U - G_\text{pH}
    \end{aligned}
\end{equation}
\begin{equation}
     \begin{aligned}
     \Delta &G_4=[E_* + 2 E_{H_2O} + 4.92 - \frac{3}{2}E_{H_2} - E_{\mathrm{OOH^*}}]\\ &\quad + [2 ZPE_{H_2O} - \frac{3}{2}ZPE_{H_2} - ZPE_{\mathrm{OOH^*}}]\\ &\quad - [2 TS_{H_2O} - \frac{3}{2}TS_{H_2} - TS_{\mathrm{OOH^*}}] \\ &\quad + G_U - G_\text{pH}
     \end{aligned} 
\end{equation}
In these expressions, $E_*$ is the total energy of the clean surface, $E_{OH^*}$, $E_{O^*}$ and $E_{OOH^*}$ are the total energies of an OH, O and OOH adsorbed on the surface. $E_{H_2O}$ and $E_{H_2}$ are the total energies of water molecule and hydrogen, respectively. ZPE is the zero point energy and TS is the temperature-entropy product. Each $\Delta G$ is calculated according to the preceding step. The entire water oxidation reaction has a $\Delta G_0$ of 4.92 eV, which is added to conclude the thermodynamic cycle of the OER. The maximum value among $\Delta G_1$ to $\Delta G_4$ is defined as $\Delta G_{max}$, which reflects the energy barrier of the complete OER pathway and is the main criterion to compare the water oxidation performance in the type-II and Z-scheme processes. This comparison shows the more favourable charge transfer mechanism for the OER pathway.

Then, the photocatalytic performance of the GeC/ReS$_2$ heterostructure is studied by Z-scheme and type-II charge-transfer mechanisms. Hydrogen evolution and oxygen evolution reactions in acidic and alkaline media are shown by means of Gibbs free energy diagrams. Figs.~\ref{fig:gibbs-free-energy-Z-scheme}(a)–(c) show the HER, acidic OER and alkaline OER diagrams for the Z-scheme mechanism, and Figs.~\ref{fig:gibbs-free-energy-Type-II}(a)–(c) show the same diagrams for the type-II mechanism. In all these diagrams the red dashed line represents the border between the two states, where the "light-off" state (U=0) is above the red dashed line and the "light-on" state (U $\neq$ 0) is below the red dashed line.

\subsubsection*{Z-scheme Path}
\begin{figure*}[ht!]
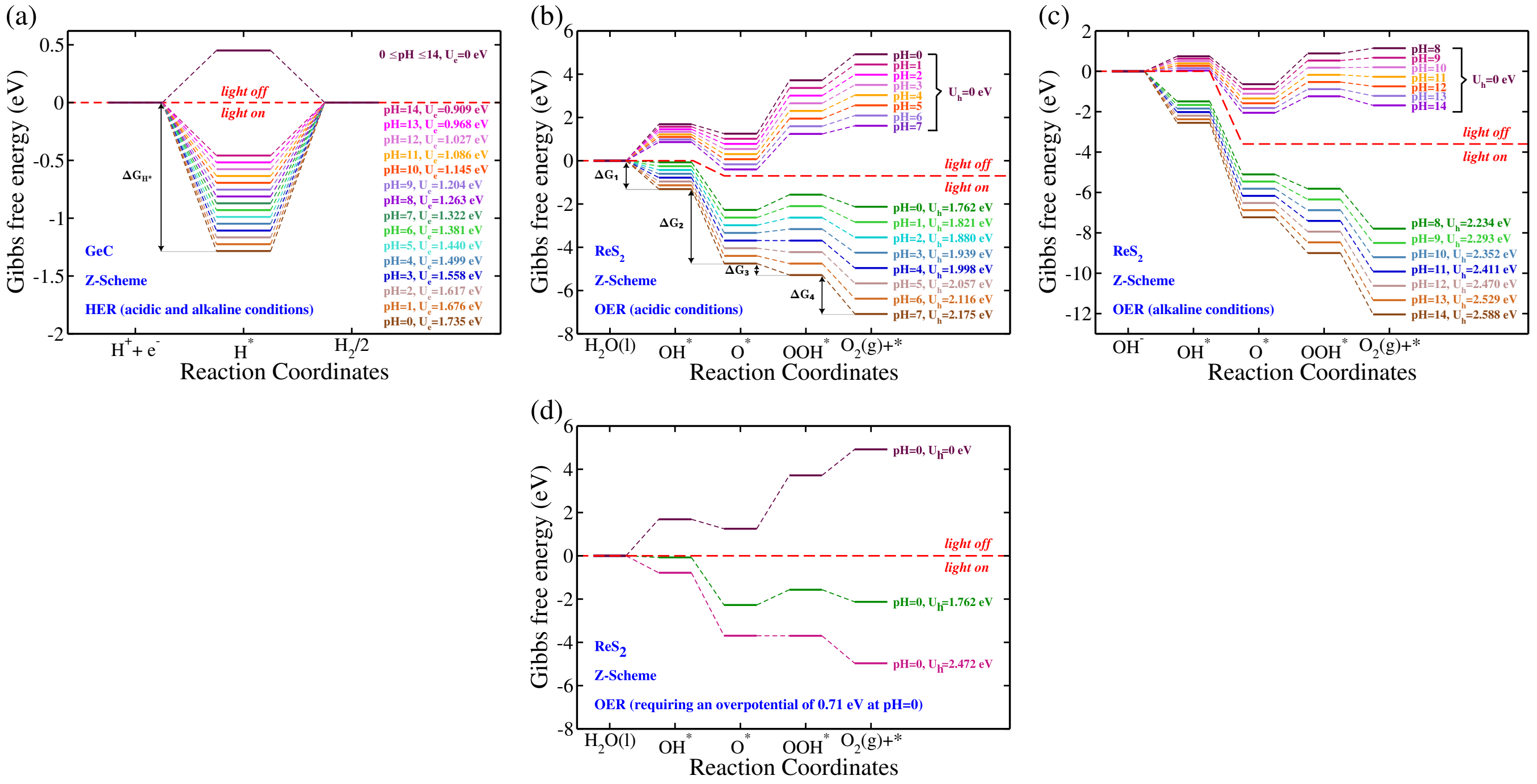

\centering
\begin{overpic}[width=\textwidth]{./HER-OER-Z-Scheme2.pdf}
\end{overpic}
\caption{Gibbs free energy diagrams for the photocatalytic water splitting pathways in the Z-scheme mechanism. (a) HER, (b) acidic OER, and (c) alkaline OER under dark and illuminated conditions at different pH values. (d) Acidic OER at pH=0 under illumination with an overpotential of 0.71 eV. Note that HER and alkaline OER are already spontaneous under illumination and require no overpotential, hence no corresponding panels are provided.}
\label{fig:gibbs-free-energy-Z-scheme}
\end{figure*}

In the Z-scheme mechanism, the HER occurs on the GeC surface and the OER on the ReS$_2$ surface. Fig.~\ref{fig:adsorption-Z-scheme} shows the most stable adsorption configurations of the H atom on the GeC surface and the OH, O, and OOH intermediates on the ReS$_2$ surface. Based on this adsorption configuration, the Gibbs free energy curve for the HER is plotted in Fig.~\ref{fig:gibbs-free-energy-Z-scheme}(a). The computed free energy for hydrogen adsorption in the light-off state at pH=0 is $\Delta G_{H^*}=0.451$ eV. This result is close to the ideal value ($\Delta G_{H^*}=0$) and suggests a good hydrogen adsorption ability of the GeC surface.

In this mechanism, the photogenerated electron potential is the difference between the CBM of the GeC layer and the hydrogen reduction potential, which is U$_e=1.735$ eV at pH=0. U$_e$ drops from 1.735 to 0.909 eV when pH increases from 0 to 14, but remains higher than the magnitude of $\Delta G_{H^*}$. Hence, HER can be driven spontaneously at all pH values (0-14) in the absence of any overpotential. Under irradiation, the value of $\Delta G_{H^*}$ varies from -1.284 eV at pH=0 to -0.458 eV at pH=14, implying that $\Delta G_{H^*}$ at higher pH is closer to zero and the hydrogen adsorption and desorption are more balanced. The complete set of HER Gibbs free energy values at various pH levels for this mechanism is detailed in Table~\ref{tab:HER-Z-scheme} of the Supporting Information.

The Gibbs free energy diagram for OER in acidic medium is presented in Fig.~\ref{fig:gibbs-free-energy-Z-scheme}(b). To explore this reaction, OH, O and OOH intermediates were adsorbed on the surface of ReS$_2$ layer. In the light-off state at pH=0, the changes of the Gibbs free energy for each intermediate state are $\Delta G_1=1.688$, $\Delta G_2=-0.438$, $\Delta G_3=2.468$, and $\Delta G_4=1.202$ eV, respectively. The third step, production of the intermediate OOH{$^*$} (2.468 eV), is the highest energy barrier and the rate-determining step in the OER route. The OER is non-spontaneous and thermodynamically unfavorable in each step in the light-off state.

In the light-on state, the potential of the photogenerated hole is calculated as the difference between the hydrogen reduction potential and the VBM of the ReS$_2$ layer, which is U$_h$=1.762 eV at pH=0. With this potential, the energy barrier of the OOH* intermediate is lowered, and the free energies of all intermediates become negative. However, in the pH range of 0-3, the free energy values of the intermediates are negative, but the reaction steps do not show a downhill trend, and the reaction cannot proceed to completion. For pH greater than 4, the OER steps are exergonic, and the reaction is spontaneous. An overpotential needs to be imposed on the system to drive the OER spontaneously in the pH range of 0-3. The biggest energy barrier of this pathway is $\Delta G_3$=2.468 eV and the photogenerated hole potential at pH=0 is U$_h=1.762$ eV. Thus, an overpotential of 0.71 eV is necessary to make OER spontaneous in the pH range of 0-3. The effect of this applied overpotential is shown in Fig.~\ref{fig:gibbs-free-energy-Z-scheme}(d), where the energy barrier of the OOH intermediate is completely eliminated, making the reaction pathway fully downhill.

Fig.~\ref{fig:gibbs-free-energy-Z-scheme}(c) shows the Gibbs free energy diagram of the OER in alkaline media. The reaction is not spontaneous for the pH range 8-14 when the system is in the light-off condition. However, the situation is significantly changed when considering the photogenerated hole potential (light-on state). In alkaline medium, U$_h$ rises from 2.234 eV at pH=8 to 2.588 eV at pH=14, which is much bigger than the energy barrier of the OOH{$^*$} intermediate. Thus, all pH levels from 8 to 14 are spontaneous upon light irradiation. These $\Delta G$ values are more negative in an alkaline medium than in an acidic medium, indicating a higher thermodynamic driving force for water oxidation in this medium. 
The detailed Gibbs free energy changes for all OER steps at different pH values in the Z-scheme mechanism are summarized in Table~\ref{tab:OER-Z-scheme} of the Supporting Information.

\subsubsection*{Type-II Path}
\begin{figure*}[ht!]
\centering
\begin{overpic}[width=\textwidth]{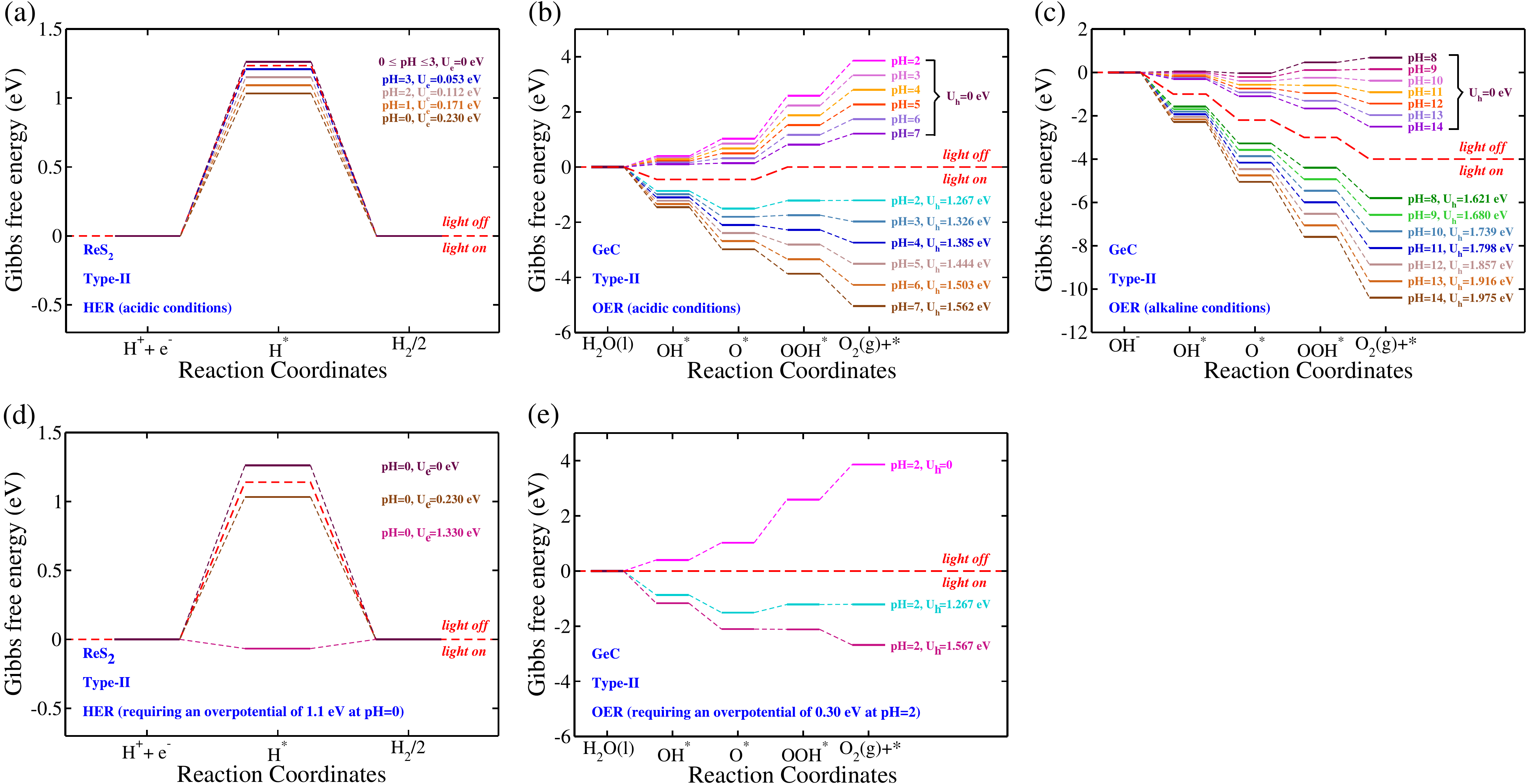}
\end{overpic}
\caption{Gibbs free energy diagrams for the photocatalytic water splitting pathways in the Type-II mechanism. (a) HER, (b) acidic OER, and (c) alkaline OER under dark and illuminated conditions at different pH values. (d) HER at pH=0 and (e) acidic OER at pH=2 under illumination with overpotentials of 1.1 eV and 0.30 eV, respectively. Note that alkaline OER (panel c) is already spontaneous under illumination and requires no overpotential, hence no corresponding panel is provided.}
\label{fig:gibbs-free-energy-Type-II}
\end{figure*}

In the type-II mechanism, the OER occurs on the GeC surface and the HER on the ReS$_2$ surface, which is different from the Z-scheme. The most stable adsorption configurations of the H atom on the ReS$_2$ surface and the OH, O and OOH intermediates on the GeC surface are shown in Fig.~\ref{fig:adsorption-Type-II}. According to this adsorption configuration, the Gibbs free energy curve for HER is shown in Fig.~\ref{fig:gibbs-free-energy-Type-II}(a). The computed free energy of hydrogen adsorption at pH=0 in the light-off state is $\Delta G_{H^*}=1.263$~eV. This number is far from the ideal value, indicating that the adsorption of hydrogen on the ReS$_2$ surface is unfavourable.

The calculated potential of the photogenerated electrons, as the difference between the CBM of the ReS$_2$ layer and the hydrogen reduction potential, is equivalent to 0.230 eV at pH=0, which is not enough to drive the reaction spontaneously. The band alignment analysis demonstrated that the position of the CBM of the ReS$_2$ layer confines the HER to the pH range of 0–3. However, U$_e$ decreases further from 0.230 eV at pH=0 to 0.053 eV at pH=3, which is insufficient to overcome the hydrogen adsorption free energy, meaning the reaction does not occur. Therefore, the HER in the type-II mechanism is not spontaneous at any of the pH values and is hence not suited for hydrogen production. In this mechanism, the hydrogen adsorption free energy is 1.263 eV, while the photogenerated electron potential at pH=0 is just 0.230 eV. Therefore, a further overpotential of 1.1 eV is needed to make the HER spontaneous at pH=0. The comprehensive pH-dependent HER energetics for the Type-II mechanism are detailed in Table~\ref{tab:HER-Type-II} of the Supporting Information. This necessary overpotential is seen in Fig.~\ref{fig:gibbs-free-energy-Type-II}(d), where the barrier is entirely removed.

As discussed before in the band alignment analysis, thermodynamic restrictions limit the OER in this type-II configuration to occur exclusively in the pH range of 2–14. Thus, the Gibbs free energy diagram for OER in an acidic medium is depicted in Fig.~\ref{fig:gibbs-free-energy-Type-II}(b). To analyze the process, OH, O, and OOH intermediates were adsorbed on the surface of the GeC layer. The Gibbs free energy changes of the four chemical steps in the light-off state at pH=2 were $\Delta G_1=0.399$, $\Delta G_2=0.627$, $\Delta G_3=1.560$ and $\Delta G_4=1.274$ eV, respectively. The third step, the formation of the OOH$^*$ intermediate (1.560 eV), is the highest energy barrier and the rate-determining step in the OER process. In the off state, the OER is thermodynamically unfavorable for all steps.

The photogenerated hole potential is given by the energy difference of the hydrogen reduction potential and the VBM of the GeC layer (1.267 eV at pH=2). Applying this potential lowers the energy barrier for the OOH$^*$ intermediate, and the $\Delta G$ values become negative. However, the reaction is not spontaneous at pH=2 and 3. For pH=4-7, the OER steps are downhill, and the reaction is spontaneous. An overpotential is necessary to drive the OER spontaneously at pH=2 and 3 under illumination. The highest energy barrier in this process is 1.560 eV, the potential of the photogenerated hole at pH=2 is 1.267 eV. Thus, an overpotential of 0.30 eV is needed. The impact of this applied overpotential is depicted in Fig.~\ref{fig:gibbs-free-energy-Type-II}(e), demonstrating a completely downhill pathway.

Fig.~\ref{fig:gibbs-free-energy-Type-II}(c) shows the Gibbs free energy profile for the OER in alkaline media. The total trend of the OER steps is not downhill in the light-off state for pH values from 8 to 10, and the reaction is thermodynamically unfavorable. However, the OER steps exhibit a downward trend from pH=11 to 14 and the reaction is spontaneous. The potential of the photogenerated hole, U$_h$, increases from 1.621 eV at pH=8 to 1.975 eV at pH=14. With this potential, when considering the full thermodynamic pathway, all pH values from 8 to 14 become spontaneous under illumination and the $\Delta G$ values are negative in all steps. The U$_h$ values of the type-II mechanism are much smaller than those of the Z-scheme mechanism. In Table~\ref{tab:OER-Type-II} of the Supporting Information, you can find the complete set of step-by-step Gibbs free energy data for OER in the Type-II mechanism for all pH values.

\subsubsection*{Determination of the Charge Transfer Mechanism}
The criterion suggested by Liu et al.~\cite{Yaozhong2024} is used to determine the leading charge transfer process in this heterostructure. Under this criterion, the pathway having a smaller maximum energy barrier between the type-II and Z-scheme processes is considered as the dominant pathway. However, Pingak et al.~\cite{Redi2025} demonstrated that this criterion alone is not sufficient, since the photogenerated charge transfer proceeds in the light-on state and not in the light-off state. Hence, the analysis must be performed in the light-on state to align with the direction of the heterostructure's built-in electric field. The pathway with the smallest (most negative) change in Gibbs free energy in the light-on state is referred to as the dominant charge transfer pathway.

The HER in the type-II mechanism does not proceed spontaneously in the light-on state and requires an overpotential of 1.1 eV. In contrast, for the Z-scheme mechanism, the HER naturally occurs in the pH range of 0 $\le$ pH $\le$ 14 without any overpotential. However, although a type-II mechanism would lead to near-thermoneutral hydrogen adsorption, it requires a considerable energy input, with an overpotential of 1.1 eV. Conversely, the Z-scheme works totally driven by solar energy. This large difference indicates that the Z-scheme mechanism is suitable for producing hydrogen in a wide range of pH values, while the type-II mechanism does not show this capability.

In contrast, for the Z-scheme mechanism of the PtSSe/phosphorene~\cite{Poonam2024} heterostructure, the Gibbs free energy of hydrogen adsorption is reported to be 1.78 eV, which decreases to 1.12 eV by applying the photogenerated electron potential and requires 1.13 eV for the reaction to be spontaneous. The photogenerated electron potential reduces the $\Delta G_{H^*}=1.76$ eV in the type-II mechanism of PtSSe/phosphorene to 1.08 eV, and a 1.09 eV overpotential is required for the reaction to be spontaneous. From the comparison, it is evident that GeC/ReS$_2$ heterostructure is better than PtSSe/phosphorene in HER in both mechanisms and works in the Z-scheme mechanism with no overpotential.

In an acidic medium and in the light-on state, a similar process is observed for both mechanisms, so that both the Z-scheme and type-II mechanisms can drive the OER spontaneously in the pH range of 4 $\le$ pH $\le$ 7. However, for the reaction to spontaneously occur, the Z-scheme requires an overpotential of 0.71 eV in the pH range of 0–3, and the type-II requires an overpotential of 0.30 eV at pH=2. In an alkaline medium, the behaviour is similar, and in the pH range of 8 $\le$ pH $\le$ 14, the reactions proceed spontaneously for both mechanisms. This result shows that the GeC/ReS$_2$ heterostructure, regardless of the charge transfer type (Z-scheme or type-II), is a suitable photocatalyst for spontaneous OER. However, it is clearly seen that at all steps of the reaction, the Z-scheme mechanism shows a lower (more negative) Gibbs free energy difference than the type-II mechanism, due to its larger photogenerated hole potential. The photogenerated hole potential in the Z-scheme is 1.762 eV at pH=0 and increases with increasing pH up to 2.588 eV at pH=14, whereas in the type-II mechanism, this value ranges from 1.267 eV at pH=2 to 1.975 eV at pH=14.

The OER performance of the GeC/ReS$_2$ heterostructure is also comparable to those reported for other heterostructures. In the InS/GaTe heterostructure~\cite{Redi2025}, the OER is spontaneous for 2 $\le$ pH $\le$ 14, while for GeC/ReS$_2$ the OER is spontaneous for 4 $\le$ pH $\le$ 14. Conversely, the AlN/PtSSe~\cite{Yaozhong2024} heterostructure can only undergo spontaneous OER at pH=7. GeC/ReS$_2$ has a larger pH range for spontaneous water oxidation compared to AlN/PtSSe, and is better in this respect. The overpotential required for spontaneous OER in GeC/ReS$_2$ is 0.71 eV at pH=0 in the Z-scheme mechanism and 0.30 eV at pH=2 in the type-II mechanism. The PtSSe/phosphorene~\cite{Poonam2024} heterostructure, however, needs an overpotential of 1.08 eV in the Z-scheme and 1.17 eV in the type-II mechanism for spontaneous OER at pH=0. Consequently, GeC/ReS$_2$ requires less overpotential than PtSSe/phosphorene in both mechanisms and demonstrates greater performance for water oxidation.

Thus, the examination of the two mechanisms under the light-on state reveals that the Z-scheme mechanism is more advantageous for the GeC/ReS$_2$ heterostructure for both the HER and OER, which is also consistent with the direction of the built-in electric field in this heterostructure.

\subsection{Optical Absorption  Properties}
\begin{figure}[ht!]
\centering
\includegraphics[scale=0.25]{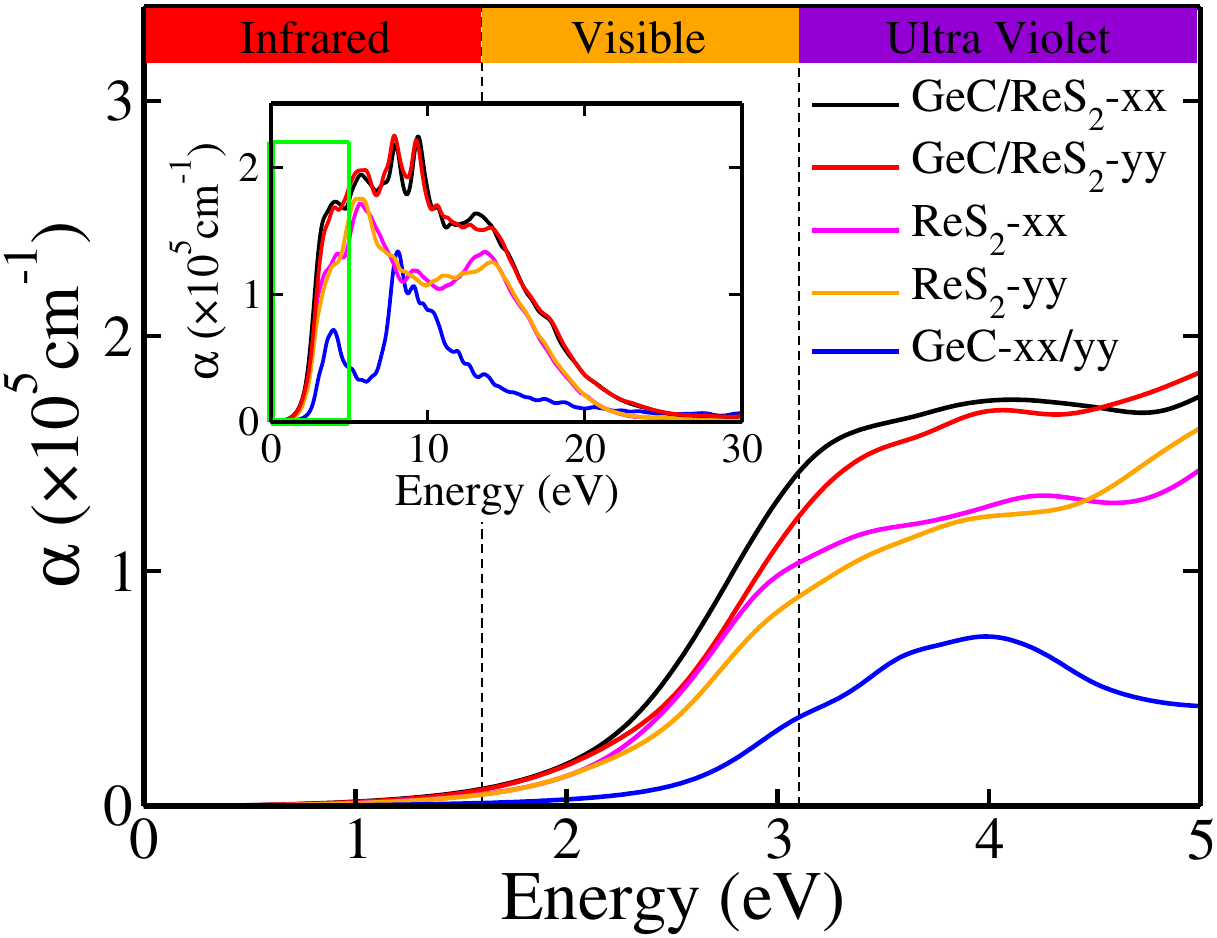}
\caption{Optical absorption spectra of the GeC and ReS$_2$ monolayers and the GeC/ReS$_2$ heterostructure.}
\label{fig:optical-absorption}
\end{figure}

The optical properties of a material are characterized by its absorption spectrum. The complex dielectric function is used to calculate the optical absorption coefficient ($\alpha(\omega)$)~\cite{Gajdo2006}:
\begin{equation}
    \alpha(\omega)=\frac{\sqrt{2}\,\omega}{c} \left( \sqrt{\varepsilon_1^2(\omega) + \varepsilon_2^2(\omega)} - \varepsilon_1(\omega) \right)^{1/2} ,
\end{equation}
where $\varepsilon_1(\omega)$ and $\varepsilon_2(\omega)$ are the real and imaginary parts of the dielectric function, and $\omega$ is the angular frequency of the photon. The absorption coefficient is an important metric for assessing the photocatalytic performance of the GeC/ReS$_2$ heterostructure and provides information on the interaction between light and matter. The formation of a greater number of electrons and holes are a direct consequence of an increased absorption intensity and a greater number of absorption peaks, thereby enhancing photocatalytic activities.
Significant absorption in the visible range is essential for the efficient harvesting of solar energy, as over 46\% of the solar spectrum is in the visible range and only 4\% is in the ultraviolet. Consequently, the absorption coefficients of the GeC monolayer, the ReS$_2$ monolayer, and the GeC/ReS$_2$ heterostructure were determined by calculation and compared in Fig.~\ref{fig:optical-absorption}. This enables us to compare the photocatalytic activity of the GeC/ReS$_2$ heterostructure to that of the constituent monolayers and to investigate the change in light absorption.

The absorption spectrum of the GeC monolayer is identical in the xx and yy directions. The absorption is in the low range throughout the spectrum. Nevertheless, the ReS$_2$ monolayer's absorption is distinct in the xx and yy directions as a result of its structural anisotropy. The absorption of this ReS$_2$ monolayer is greater than that of the GeC monolayer in both directions throughout the entire spectral range. The GeC/ReS$_2$ heterostructure also exhibits anisotropic absorption. It is intriguing that the absorption of the GeC/ReS$_2$ heterostructure is greater than that of the ReS$_2$ and GeC monolayers throughout the entire spectrum, particularly in the visible region. This is because the band gap of the GeC/ReS$_2$ heterostructure is narrower than that of the constituent monolayers, which enables electronic transitions at lower photon energy. Consequently, the absorption is increased.

Consequently, the photocatalytic activity of the GeC/ReS$_2$ heterostructure is improved in comparison to that of the individual monolayers due to the improved light absorption. The GeC/ReS$_2$ heterostructure is a feasible candidate for overall water splitting due to the enhanced light absorption and the optimal Z-scheme charge transfer pathway.

\subsection{Strain-Induced Modulation of Electronic and Optical Properties}
The electrical and optical properties of semiconductors can be effectively modulated through strain engineering. In particular, low-dimensional materials may be subjected to substantially greater strains than their bulk counterparts. The electronic band gap of the GeC/ReS$_2$ heterostructure was investigated in relation to in-plane strain.

The biaxial strain modifies the lattice constant, which is determined as~\cite{Karbasizadeh2022,Deming2024}:
\begin{equation}
    \varepsilon=\frac{a_{\text{strained}} - a_{\text{relaxed}}}{a_{\text{relaxed}}} \times 100\% ,
\end{equation}
where $a_{\text{strained}}$ is the lattice constant under biaxial strain and $a_{\text{relaxed}}$ is the equilibrium lattice constant of the GeC/ReS$_2$ heterostructure, and $\varepsilon$ denotes the applied strain percentage. The electrical and optical characteristics of the GeC/ReS$_2$ heterostructure can be optimized through the use of biaxial strain, which is advantageous for customizing its water splitting capabilities.

\begin{figure}[ht!]
\centering
\begin{overpic}[scale=0.22]{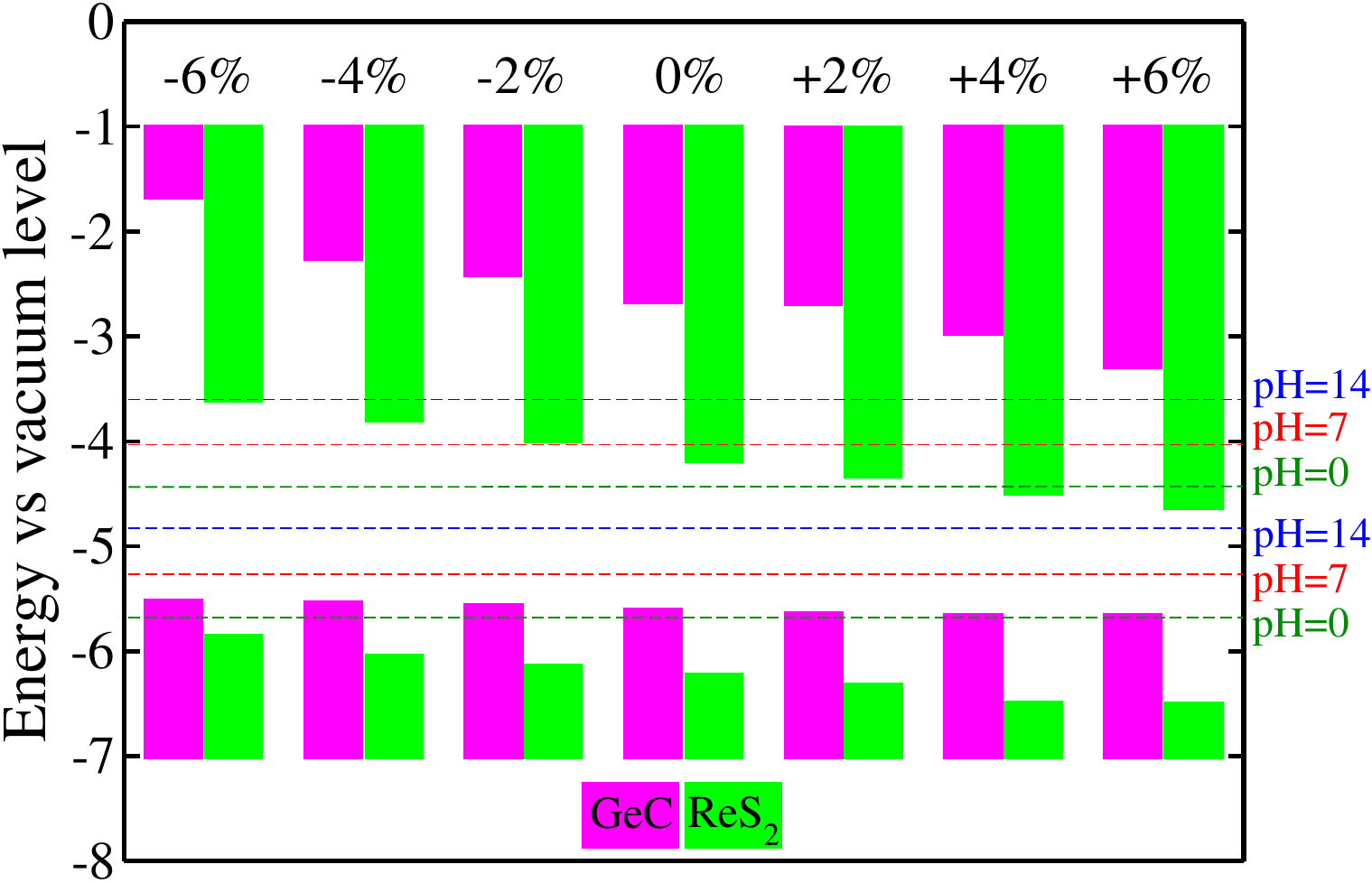}
\end{overpic}
\caption{Band alignment of the GeC/ReS$_2$ heterostructure under biaxial strain from -6\% to +6\%.}
\label{fig:bandalignment-strain}
\end{figure}

Band alignment diagrams (Fig.~\ref{fig:bandalignment-strain}) and band structures (Fig.~\ref{fig:bands-strain}) illustrate that the GeC/ReS$_2$ heterostructure exhibits a staggered band alignment for all applied strains from -6\% to +6\%. The heterostructure maintains a staggered band alignment, which is suitable for water splitting in a variety of pH conditions, despite the fact that the band gap increases with compressive strain and decreases with tensile strain. In the unstrained GeC/ReS$_2$ heterostructure, the Z-scheme pathway is the predominant charge transfer mechanism, as detailed in the preceding section. The spatial separation of the band boundaries and the direction of the built-in electric field remain unaltered within the applied strain range. Furthermore, the average electrostatic potential diagrams for the various strain levels, as illustrated in Fig.~\ref{fig:potentials-strain} in the Supporting Information, further substantiate the preservation of this built-in electric field direction. Consequently, the Z-scheme's charge transfer mechanism is preserved. This illustrates the potential of an adjustable band structure for photocatalytic applications.

\begin{figure}[ht!]
\centering
\includegraphics[scale=0.25]{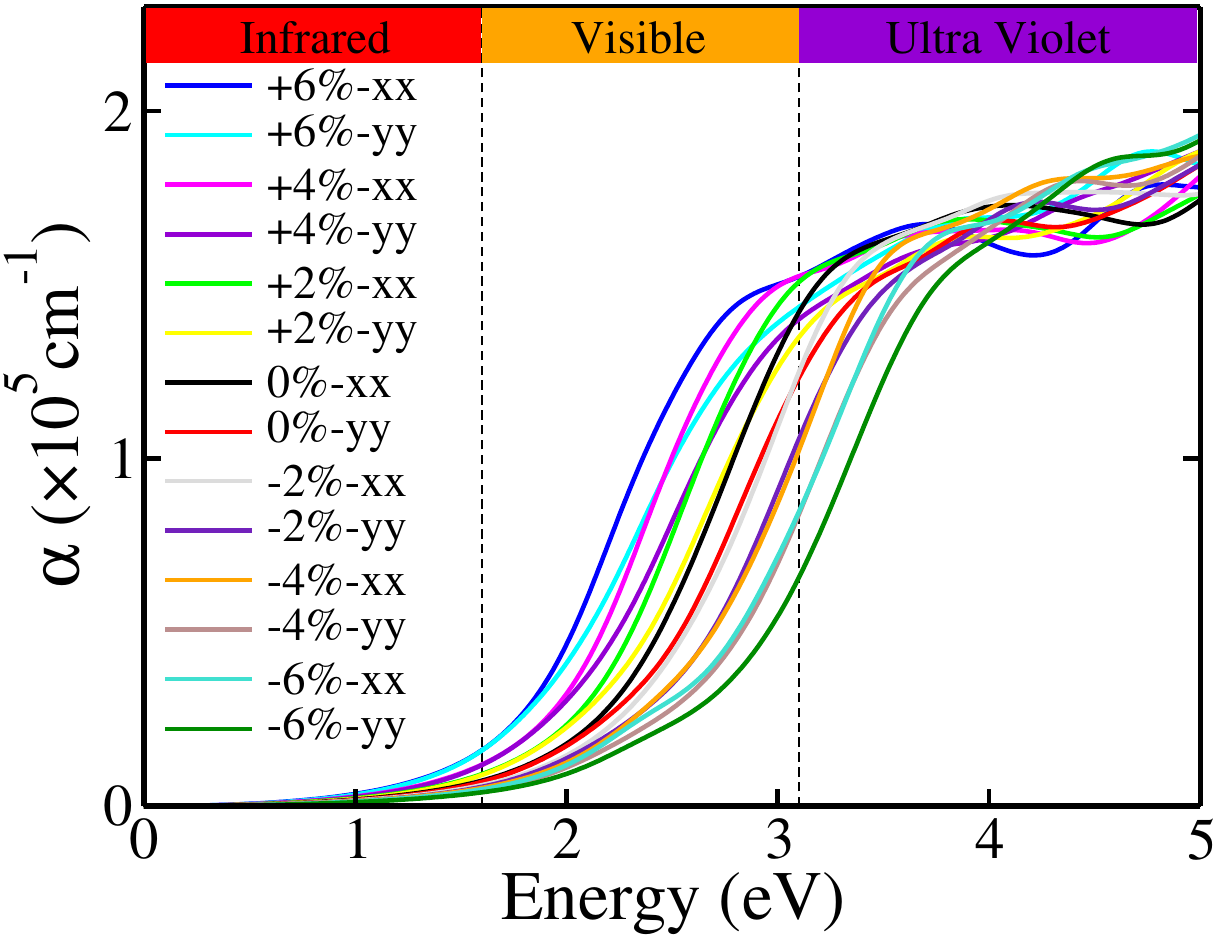}
\caption{Optical absorption spectra of the GeC/ReS$_2$ heterostructure under biaxial strain from -6\% to +6\%.}
\label{fig:absorption-strain}
\end{figure}

Additionally, strain engineering can be employed to adjust the optical properties of the GeC/ReS$_2$ heterostructure. The utilization of solar energy is enhanced by the tuning of the absorption spectra to the visible region through the use of in-plane strain. In this regard, we determined the absorption coefficients of the GeC/ReS$_2$ heterostructure under biaxial strains ranging from -6\% to +6\% and presented the results in Fig.~\ref{fig:absorption-strain}. These findings indicate that the absorption of visible light can be substantially increased by tensile strain in comparison to the unstrained state.

\subsection{Solar to Hydrogen (STH)}

Solar-to-hydrogen (STH) efficiency ($\eta_{STH}$) is an important parameter to evaluate the practical performance of a photocatalyst. It is defined as a product of light absorption efficiency ($\eta_{abs}$) and carrier usage efficiency ($\eta_{cu}$), as:
\begin{equation}
    \eta_{STH}=\eta_{abs} \times \eta_{cu}
\end{equation}
The $\eta_{abs}$ and $\eta_{cu}$ are calculated by the following equations:
\begin{equation}
    \eta_{abs}=\frac{\int_{E_g}^{\infty} P(\hbar\omega)\, d(\hbar\omega)}{\int_{0}^{\infty} P(\hbar\omega)d(\hbar\omega)}\,
\end{equation}
\begin{equation}
    \eta_{cu}=\frac{\Delta G \int_{E}^{\infty} \frac{P(\hbar\omega)}{\hbar\omega} \, d(\hbar\omega)}{\int_{E_g}^{\infty} P(\hbar\omega) \, d(\hbar\omega)}
\end{equation}
where $P(\hbar\omega)$ is the AM1.5G solar spectrum flux at photon energy $\hbar\omega$, and $\Delta G=1.23$~eV is the redox potential for water splitting. The minimum photon energy for the redox processes ($E$) is calculated from the overpotentials of the hydrogen evolution reaction ($\chi(H_2)$) and the oxygen evolution reaction ($\chi(O_2)$) as follows:
{\small
\begin{equation}
    E=
    \begin{cases}
      E_g & \chi(H_2) \ge 0.2, \, \chi(O_2) \ge 0.6 \\
       E_g + 0.2 - \chi(H_2) & \chi(H_2) < 0.2, \, \chi(O_2) \ge 0.6 \\
       E_g + 0.6 - \chi(O_2) & \chi(H_2) \ge 0.2, \, \chi(O_2) < 0.6 \\
       E_g + 0.8 - \chi(H_2) - \chi(O_2) & \chi(H_2) < 0.2, \, \chi(O_2) < 0.6 
    \end{cases}
\end{equation}
}
Here $\chi(H_2)$ is the difference in energy between the conduction band minimum and the hydrogen reduction potential and $\chi(O_2)$ is the difference in energy between the water oxidation potential and the valence band maximum. The STH efficiency is pH-dependent because the overpotentials $\chi(H_2)$ and $\chi(O_2)$ vary with pH. The computed overpotentials for the GeC/ReS$_2$ heterostructure at pH=0 are $\chi(H_2)=1.735$~eV ($\ge 0.2$ eV) and $\chi(O_2)=0.532$ eV ($< 0.6$ eV). The band gap is $E_g=1.38$ eV. Thus the third condition is applicable and $E=1.38+0.6-0.532=1.447$ eV.

Based on these characteristics, we computed the STH efficiency of the pristine GeC/ReS$_2$ heterostructure. The efficiency increases gradually as the pH increases. It begins at 38.02\% at pH=0, increases to 40.10\% at pH=1 and remains at 41.40\% for pH=2–14, which is far higher than the 10\% economic viability threshold, indicating its promising practical potential.

To further enhance the performance, the influence of biaxial strain on the STH efficiency was studied. The modulation of the STH efficiency is consistent with the above-mentioned optical absorption measurements. The application of tensile strain throughout the full pH range significantly improves the efficiency. The efficiency at pH=0 increases to 45.14\% under +2\% strain and saturates at 46.78\% at pH $\geq 1$. Increasing the tensile strain further to +4\% and +6\% results in a monotonic increase, where the STH efficiencies are substantially boosted to 54.24\% and 60.68\% over all pH values, respectively. The large enhancement is due to the reduced band gap and the enhanced visible-light absorption generated by tensile strain.

In contrast, compressive strain causes a significant decrease in efficiency. At a strain of -2\%, the efficiency is 28.36\% for pH=0 and remains constant at 34.42\% at pH $\ge 3$. As the compressive strain increases to -4\% and -6\%, the efficiency further decreases. For instance, under -6\% strain, the STH efficiency reaches a maximum of 22.46\% in the alkaline range (pH 8–14) and drops to a minimum of 11.20\% at pH=0.

For comparison, the STH efficiency of the GeC/ReS$_2$ heterostructure is compared with other photocatalysts published in the literature to provide a perspective on the above results. The pristine heterostructure shows an efficiency of 41.40\% which is better than some of the reported systems, such as SiC/PtSe$_2$ (34.7\%)~\cite{LI2024}, GeS/GeSe (31.21\%)~\cite{XIE2024}, g-C$_6$N$_6$/InP (27.32\%)~\cite{Wenna_2022}, GeH/InSe (24.72\%)~\cite{LI2024-2}, and PtSSe/$\gamma$-phosphorene (24.86\%)~\cite{Chauhan2024}. Significantly, a tensile strain of +6\% is applied to enhance the efficiency to 60.68\%. This value is also larger than that of InS/GaTe heterostructure (44.8\%)~\cite{Redi2025} and the most reported theoretical value for MoTe$_2$/BAs (56.32\%)~\cite{CAO2021-2}. In general, the strain engineered GeC/ReS$_2$ heterostructure demonstrates a high potential for practical solar water splitting far beyond the 10\% barrier for economic feasibility.

The maximum STH efficiency of 60.68\% is achieved under +6\% tensile strain, indicating that tensile strain engineering is a promising approach to maximize solar energy conversion in the GeC/ReS$_2$ heterostructure. The STH efficiency is thoroughly examined in Table~\ref{tab:STH} of the Supporting Information for all strain levels and pH conditions.

%\subsection{Ehrenfest dynamics with or without excited carrier using real-time TDDFT.}

\section{Conclusion}\label{sec:conclusion}
The GeC/ReS$_2$ heterostructure's optical performance, photocatalytic efficiency, electronic properties, and structural stability are investigated through first-principles calculations. Its thermodynamic, dynamic, and thermal stability, as well as a staggered band alignment and an indirect band gap of 1.38 eV, were verified by the analysis. According to the interfacial charge transfer analysis, the spatial separation of the photogenerated carriers is facilitated by the built-in electric field from the GeC to the ReS$_2$ layer. By evaluating the Gibbs free energy pathways under irradiation, the direct Z-scheme mechanism was determined to be the most advantageous charge transfer pathway. The direction of the built-in electric field is consistent with the Gibbs free energy results, which further establishes the heterostructure's Z-scheme nature.

The OER can be spontaneously driven under irradiation in the pH range of $4 \le \text{pH} \le 14$ by both type-II and Z-scheme configurations. However, the Z-scheme pathway has a lower energy barrier. In addition, the type-II pathway is unable to spontaneously drive HER without an additional overpotential, whereas the Z-scheme mechanism can drive HER across the entire pH range ($0 \le \text{pH} \le 14$). At pH$\ge$2, the pristine heterostructure achieves a solar-to-hydrogen efficiency of 41.40\%, which surpasses the economic viability threshold of 10\%. Furthermore, a STH efficiency of 60.68\% is achieved by optimizing the band gap and light absorption through the application of a tensile strain of +6\%. The results indicate that the GeC/ReS$_2$ heterostructure, which has a built-in electric field and a favourable band alignment, is a prospective photocatalyst for overall water splitting.

\acknowledgements
We extend our gratitude to M. Alaei and H. Vahid for their contributions to the phonon calculation and AIMD calculation. This work is based upon research funded by the Iranian National Science Foundation (INSF) under Project No. 4039081.

\bibliography{ReS2-GeC}

\appendix  % or \appendices for IEEE
\renewcommand{\thesection}{S\arabic{section}}
\renewcommand{\thefigure}{S\arabic{figure}}
\renewcommand{\thetable}{S\arabic{table}}
\renewcommand{\theequation}{S\arabic{equation}}
\setcounter{section}{0}
\setcounter{figure}{0}
\setcounter{table}{0}
\setcounter{equation}{0}
\section{Supplementary Material}
\label{sec:supp}

\subsection{Additional Results}

\begin{figure*}[ht!]
\centering
\begin{overpic}[scale=0.35]{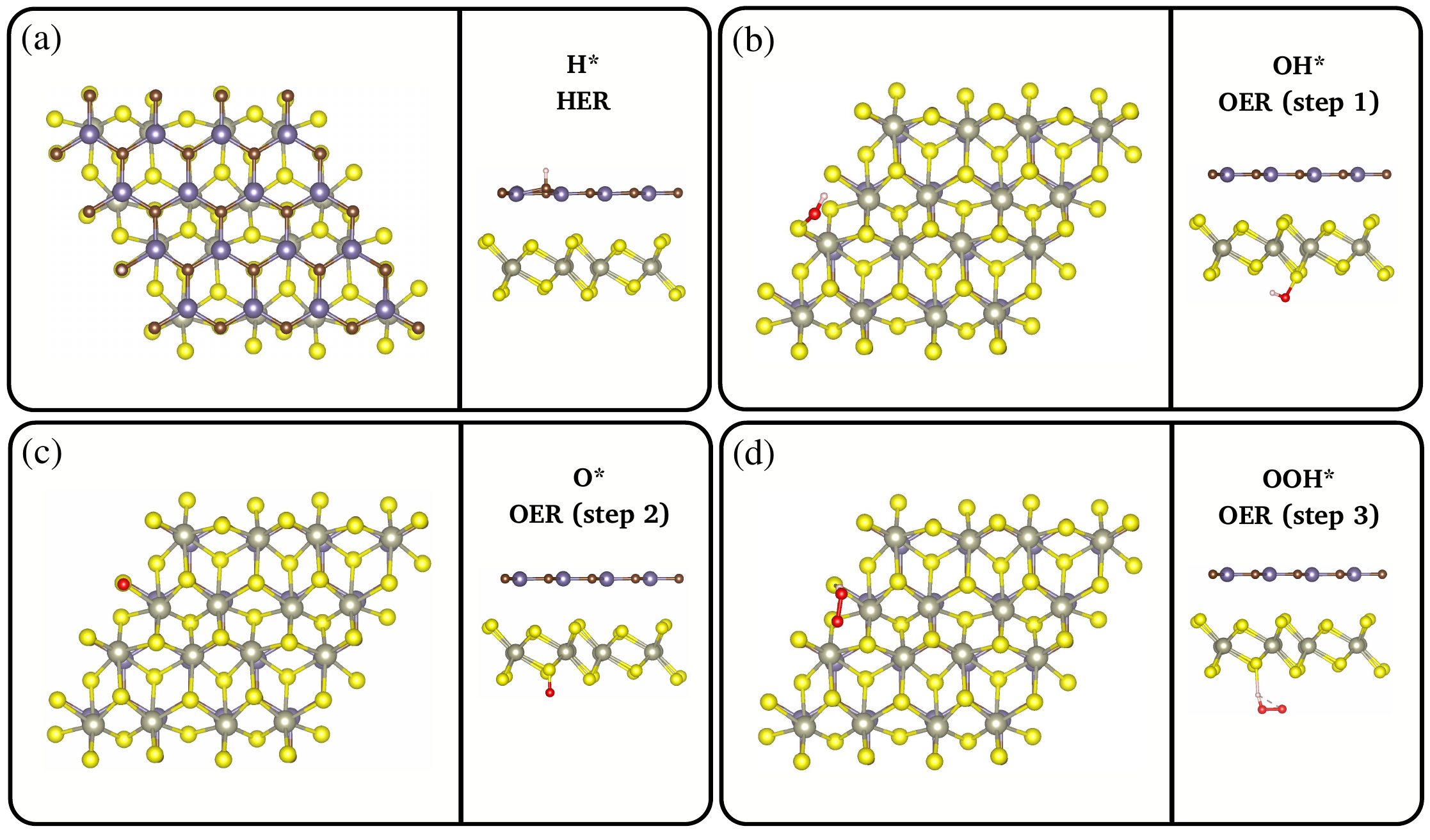}
\end{overpic}
\caption{Top and side views of the most stable adsorption configurations for the intermediates in the Z-scheme pathway: (a) H$^*$ on the GeC surface, and (b) OH$^*$, (c) O$^*$, and (d) OOH$^*$ on the ReS$_2$ surface.}
\label{fig:adsorption-Z-scheme}
\end{figure*}

\begin{figure*}[ht!]
\centering
\begin{overpic}[scale=0.35]{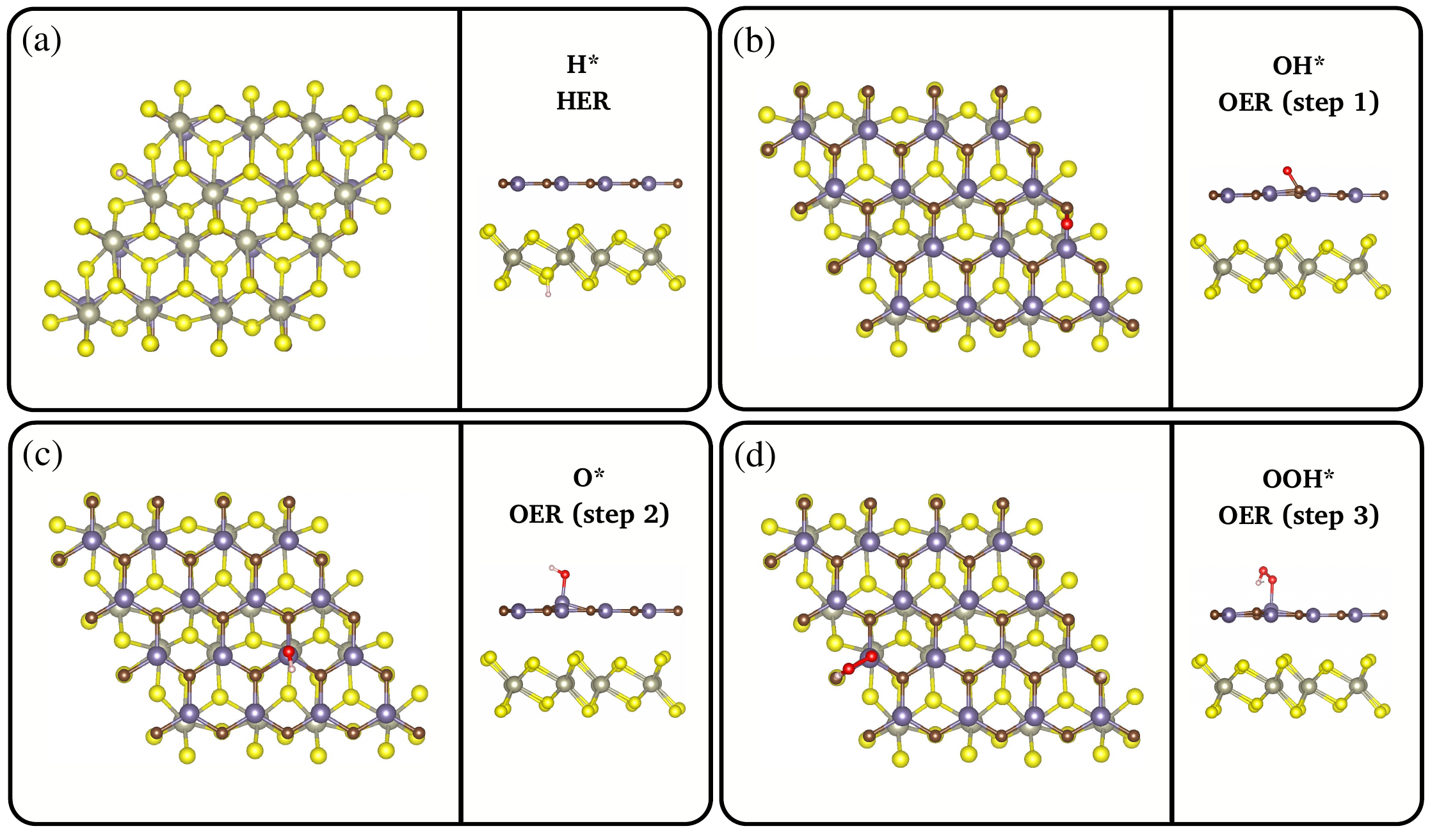}
\end{overpic}
\caption{Top and side views of the most stable adsorption configurations for the intermediates in the type-II pathway: (a) H$^*$ on the ReS$_2$ surface, and (b) OH$^*$, (c) O$^*$, and (d) OOH$^*$ on the GeC surface.}
\label{fig:adsorption-Type-II}
\end{figure*}

\begin{table*}
\caption{\label{tab:HER-Z-scheme}
Gibbs free energy of hydrogen adsorption on the GeC surface in the Z-scheme 
mechanism at different pH values. $U_e$ is the photogenerated electron 
potential and $\Delta G_{H^*}$ is the hydrogen adsorption free energy. 
Light-off ($U_e=0$) and light-on values are paired for each pH so that 
$\Delta G_{H^*}^{\text{on}}=\Delta G_{H^*}^{\text{off}} - U_e$ can be checked 
directly. The reduction potential of H\textsuperscript{+}/H\textsubscript{2} 
was computed as $E_{\text{red}}=-4.44 + 0.059 \times \text{pH}$ and 
$U_e=E_{\text{CBM}} - E_{\text{red}}$, where $E_{\text{CBM}}$ is the 
conduction-band minimum of GeC. All energies are in eV at $T=298.15$~K.}
\setlength{\tabcolsep}{12pt}
\normalsize
\begin{tabular}{c|rr@{\hspace{6pt}}|@{\hspace{6pt}}rr}
\hline
\hline
 & \multicolumn{2}{c@{\hspace{6pt}}|@{\hspace{6pt}}}{Light-off} & \multicolumn{2}{c}{Light-on} \\
pH & $U_e$ & $\Delta G_{H^*}$ & $U_e$ & $\Delta G_{H^*}$ \\
\hline
\rule{0pt}{11pt}0  & $0$ & $0.451$ & $1.735$ & $-1.284$ \\
1  & $0$ & $0.451$ & $1.676$ & $-1.225$ \\
2  & $0$ & $0.451$ & $1.617$ & $-1.166$ \\
3  & $0$ & $0.451$ & $1.558$ & $-1.107$ \\
4  & $0$ & $0.451$ & $1.499$ & $-1.048$ \\
5  & $0$ & $0.451$ & $1.440$ & $-0.989$ \\
6  & $0$ & $0.451$ & $1.381$ & $-0.930$ \\
7  & $0$ & $0.451$ & $1.322$ & $-0.871$ \\
8  & $0$ & $0.451$ & $1.263$ & $-0.812$ \\
9  & $0$ & $0.451$ & $1.204$ & $-0.753$ \\
10 & $0$ & $0.451$ & $1.145$ & $-0.694$ \\
11 & $0$ & $0.451$ & $1.086$ & $-0.635$ \\
12 & $0$ & $0.451$ & $1.027$ & $-0.576$ \\
13 & $0$ & $0.451$ & $0.968$ & $-0.517$ \\
14 & $0$ & $0.451$ & $0.909$ & $-0.458$ \\
\hline
\hline
\end{tabular}
\end{table*}

\begin{table*}
\caption{\label{tab:OER-Z-scheme}
Gibbs free energy changes for each intermediate state ($\Delta G_1$--$\Delta G_4$) and the maximum step barrier ($\Delta G_{\max}$) in the Z-scheme mechanism on the ReS\textsubscript{2} surface at different pH values. Each $\Delta G_i$ value represents the energy change of that specific step. In the light-off state, $U_h=0$ for all pH values; in the light-on state, $\Delta G_i^{\text{on}}=\Delta G_i^{\text{off}} - U_h$. The oxidation potential of H\textsubscript{2}O/O\textsubscript{2} was computed as $E_{\text{ox}}=-5.67 + 0.059 \times \text{pH}$ and $U_h=E_{\text{ox}} - E_{\text{VBM}}$, where $E_{\text{VBM}}$ is the valence-band maximum of ReS\textsubscript{2}. All energies are in eV at $T=298.15$~K.}
\setlength{\tabcolsep}{4pt}
\normalsize
\begin{tabular}{c|rrrrrr@{\hspace{6pt}}|@{\hspace{6pt}}rrrrrr}
\hline
\hline
 & \multicolumn{6}{c@{\hspace{6pt}}|@{\hspace{6pt}}}{Light-off} & \multicolumn{6}{c}{Light-on} \\
pH & $U_h$ & $\Delta G_1$ & $\Delta G_2$ & $\Delta G_3$ & $\Delta G_4$ & $\Delta G_{\max}$
 & $U_h$ & $\Delta G_1$ & $\Delta G_2$ & $\Delta G_3$ & $\Delta G_4$ & $\Delta G_{\max}$ \\
\hline
\rule{0pt}{11pt}0  & $0$ & $\phantom{-}1.688$ & $-0.438$ & $\phantom{-}2.468$ & $\phantom{-}1.202$ & $\phantom{-}2.468$ & $1.762$ & $-0.074$ & $-2.200$ & $0.706$ & $-0.560$ & $0.706$ \\
1  & $0$ & $1.629$ & $-0.497$ & $2.409$ & $1.143$ & $2.409$ & $1.821$ & $-0.192$ & $-2.318$ & $0.588$ & $-0.678$ & $0.588$ \\
2  & $0$ & $1.570$ & $-0.556$ & $2.350$ & $1.084$ & $2.350$ & $1.880$ & $-0.310$ & $-2.436$ & $0.470$ & $-0.796$ & $0.470$ \\
3  & $0$ & $1.511$ & $-0.615$ & $2.291$ & $1.025$ & $2.291$ & $1.939$ & $-0.428$ & $-2.554$ & $0.352$ & $-0.914$ & $0.352$ \\
4  & $0$ & $1.452$ & $-0.674$ & $2.232$ & $0.966$ & $2.232$ & $1.998$ & $-0.546$ & $-2.672$ & $0.234$ & $-1.032$ & $0.234$ \\
5  & $0$ & $1.393$ & $-0.733$ & $2.173$ & $0.907$ & $2.173$ & $2.057$ & $-0.664$ & $-2.790$ & $0.116$ & $-1.150$ & $0.116$ \\
6  & $0$ & $1.334$ & $-0.792$ & $2.114$ & $0.848$ & $2.114$ & $2.116$ & $-0.782$ & $-2.908$ & $-0.002$ & $-1.268$ & $-0.002$ \\
7  & $0$ & $1.275$ & $-0.851$ & $2.055$ & $0.789$ & $2.055$ & $2.175$ & $-0.900$ & $-3.026$ & $-0.120$ & $-1.386$ & $-0.120$ \\
8  & $0$ & $1.216$ & $-0.910$ & $1.996$ & $0.730$ & $1.996$ & $2.234$ & $-1.018$ & $-3.144$ & $-0.238$ & $-1.504$ & $-0.238$ \\
9  & $0$ & $1.157$ & $-0.969$ & $1.937$ & $0.671$ & $1.937$ & $2.293$ & $-1.136$ & $-3.262$ & $-0.356$ & $-1.622$ & $-0.356$ \\
10 & $0$ & $1.098$ & $-1.028$ & $1.878$ & $0.612$ & $1.878$ & $2.352$ & $-1.254$ & $-3.380$ & $-0.474$ & $-1.740$ & $-0.474$ \\
11 & $0$ & $1.039$ & $-1.087$ & $1.819$ & $0.553$ & $1.819$ & $2.411$ & $-1.372$ & $-3.498$ & $-0.592$ & $-1.858$ & $-0.592$ \\
12 & $0$ & $0.980$ & $-1.146$ & $1.760$ & $0.494$ & $1.760$ & $2.470$ & $-1.490$ & $-3.616$ & $-0.710$ & $-1.976$ & $-0.710$ \\
13 & $0$ & $0.921$ & $-1.205$ & $1.701$ & $0.435$ & $1.701$ & $2.529$ & $-1.608$ & $-3.734$ & $-0.828$ & $-2.094$ & $-0.828$ \\
14 & $0$ & $0.862$ & $-1.264$ & $1.642$ & $0.376$ & $1.642$ & $2.588$ & $-1.726$ & $-3.852$ & $-0.946$ & $-2.212$ & $-0.946$ \\
\hline
\hline
\end{tabular}
\end{table*}

\begin{table*}
\caption{\label{tab:HER-Type-II}
Gibbs free energy of hydrogen adsorption on the ReS\textsubscript{2} surface 
in the type-II mechanism at different pH values. $U_e$ is the photogenerated 
electron potential and $\Delta G_{H^*}$ is the hydrogen adsorption free 
energy. Light-off ($U_e=0$) and light-on values are paired for each pH. The 
reduction potential of H\textsuperscript{+}/H\textsubscript{2} was computed as 
$E_{\text{red}}=-4.44 + 0.059 \times \text{pH}$ and 
$U_e=E_{\text{CBM}} - E_{\text{red}}$, where $E_{\text{CBM}}$ is the 
conduction-band minimum of ReS\textsubscript{2}. All energies are in eV at 
$T=298.15$~K.}
\setlength{\tabcolsep}{12pt}
\normalsize
\begin{tabular}{c|rr@{\hspace{6pt}}|@{\hspace{6pt}}rr}
\hline
\hline
 & \multicolumn{2}{c@{\hspace{6pt}}|@{\hspace{6pt}}}{Light-off} & \multicolumn{2}{c}{Light-on} \\
pH & $U_e$ & $\Delta G_{H^*}$ & $U_e$ & $\Delta G_{H^*}$ \\
\hline
\rule{0pt}{11pt}0 & $0$ & $1.263$ & $0.230$ & $1.033$ \\
1 & $0$ & $1.263$ & $0.171$ & $1.092$ \\
2 & $0$ & $1.263$ & $0.112$ & $1.151$ \\
3 & $0$ & $1.263$ & $0.053$ & $1.210$ \\
\hline
\hline
\end{tabular}
\end{table*}

\begin{table*}
\caption{\label{tab:OER-Type-II}
Gibbs free energy changes for each intermediate state ($\Delta G_1$--$\Delta G_4$) and the maximum step barrier ($\Delta G_{\max}$) in the type-II mechanism on the GeC surface at different pH values. Each $\Delta G_i$ value represents the energy change of that specific step. In the light-off state, $U_h=0$ for all pH values; in the light-on state, $\Delta G_i^{\text{on}}=\Delta G_i^{\text{off}} - U_h$. The oxidation potential of H\textsubscript{2}O/O\textsubscript{2} was computed as $E_{\text{ox}}=-5.67 + 0.059 \times \text{pH}$ and $U_h=E_{\text{ox}} - E_{\text{VBM}}$, where $E_{\text{VBM}}$ is the valence-band maximum of GeC. All energies are in eV at $T=298.15$~K.}
\setlength{\tabcolsep}{4pt}
\normalsize
\begin{tabular}{c|rrrrrr@{\hspace{6pt}}|@{\hspace{6pt}}rrrrrr}
\hline
\hline
 & \multicolumn{6}{c@{\hspace{6pt}}|@{\hspace{6pt}}}{Light-off} & \multicolumn{6}{c}{Light-on} \\
pH & $U_h$ & $\Delta G_1$ & $\Delta G_2$ & $\Delta G_3$ & $\Delta G_4$ & $\Delta G_{\max}$
 & $U_h$ & $\Delta G_1$ & $\Delta G_2$ & $\Delta G_3$ & $\Delta G_4$ & $\Delta G_{\max}$ \\
\hline
\rule{0pt}{11pt}2  & $0$ & $0.399$ & $0.627$ & $1.560$ & $1.274$ & $1.560$ & $1.267$ & $-0.868$ & $-0.640$ & $0.293$ & $0.007$ & $0.293$ \\
3  & $0$ & $0.340$ & $0.509$ & $1.383$ & $1.098$ & $1.383$ & $1.326$ & $-0.986$ & $-0.817$ & $0.057$ & $-0.228$ & $0.057$ \\
4  & $0$ & $0.281$ & $0.391$ & $1.206$ & $0.922$ & $1.206$ & $1.385$ & $-1.104$ & $-0.994$ & $-0.179$ & $-0.463$ & $-0.179$ \\
5  & $0$ & $0.222$ & $0.273$ & $1.029$ & $0.746$ & $1.029$ & $1.444$ & $-1.222$ & $-1.171$ & $-0.415$ & $-0.698$ & $-0.415$ \\
6  & $0$ & $0.163$ & $0.155$ & $0.852$ & $0.570$ & $0.852$ & $1.503$ & $-1.340$ & $-1.348$ & $-0.651$ & $-0.933$ & $-0.651$ \\
7  & $0$ & $0.104$ & $0.037$ & $0.675$ & $0.394$ & $0.675$ & $1.562$ & $-1.458$ & $-1.525$ & $-0.887$ & $-1.168$ & $-0.887$ \\
8  & $0$ & $0.045$ & $-0.081$ & $0.498$ & $0.218$ & $0.498$ & $1.621$ & $-1.576$ & $-1.702$ & $-1.123$ & $-1.403$ & $-1.123$ \\
9  & $0$ & $-0.014$ & $-0.199$ & $0.321$ & $0.042$ & $0.321$ & $1.680$ & $-1.694$ & $-1.879$ & $-1.359$ & $-1.638$ & $-1.359$ \\
10 & $0$ & $-0.073$ & $-0.317$ & $0.144$ & $-0.134$ & $0.144$ & $1.739$ & $-1.812$ & $-2.056$ & $-1.595$ & $-1.873$ & $-1.595$ \\
11 & $0$ & $-0.132$ & $-0.435$ & $-0.033$ & $-0.310$ & $-0.033$ & $1.798$ & $-1.930$ & $-2.233$ & $-1.831$ & $-2.108$ & $-1.831$ \\
12 & $0$ & $-0.191$ & $-0.553$ & $-0.210$ & $-0.486$ & $-0.191$ & $1.857$ & $-2.048$ & $-2.410$ & $-2.067$ & $-2.343$ & $-2.067$ \\
13 & $0$ & $-0.250$ & $-0.671$ & $-0.387$ & $-0.662$ & $-0.250$ & $1.916$ & $-2.166$ & $-2.587$ & $-2.303$ & $-2.578$ & $-2.303$ \\
14 & $0$ & $-0.309$ & $-0.789$ & $-0.564$ & $-0.838$ & $-0.309$ & $1.975$ & $-2.284$ & $-2.764$ & $-2.539$ & $-2.813$ & $-2.539$ \\
\hline
\hline
\end{tabular}
\end{table*}

\begin{figure*}[ht!]
\centering
\begin{overpic}[scale=0.25]{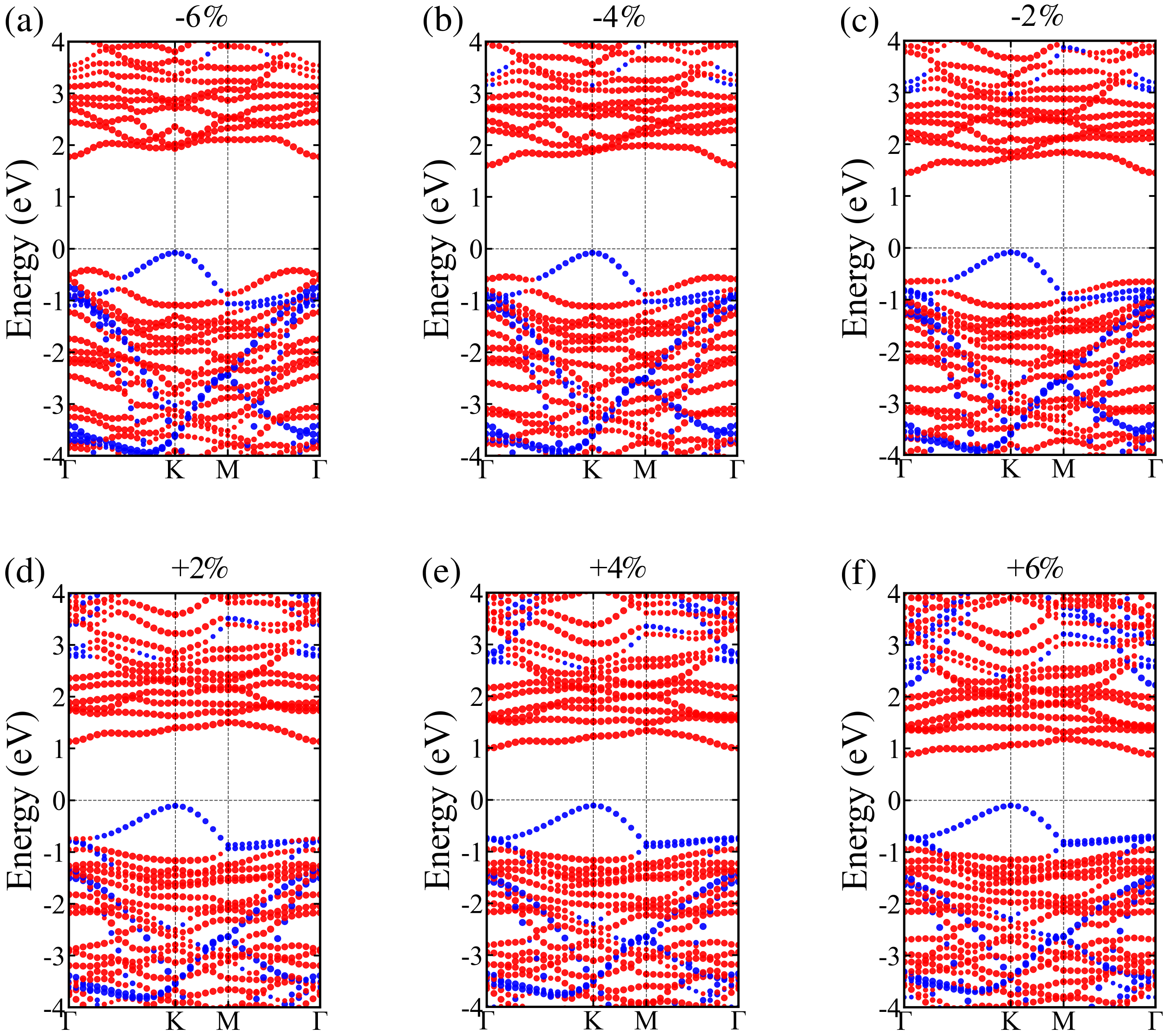}
\end{overpic}
\caption{Band structures of the GeC/ReS$_2$ heterostructure under biaxial strain ranging from -6\% to +6\%: (a) -6\%, (b) -4\%, (c) -2\%, (d) +2\%, (e) +4\%, and (f) +6\%.}
\label{fig:bands-strain}
\end{figure*}

\begin{figure*}[ht!]
\centering
\begin{overpic}[scale=0.25]{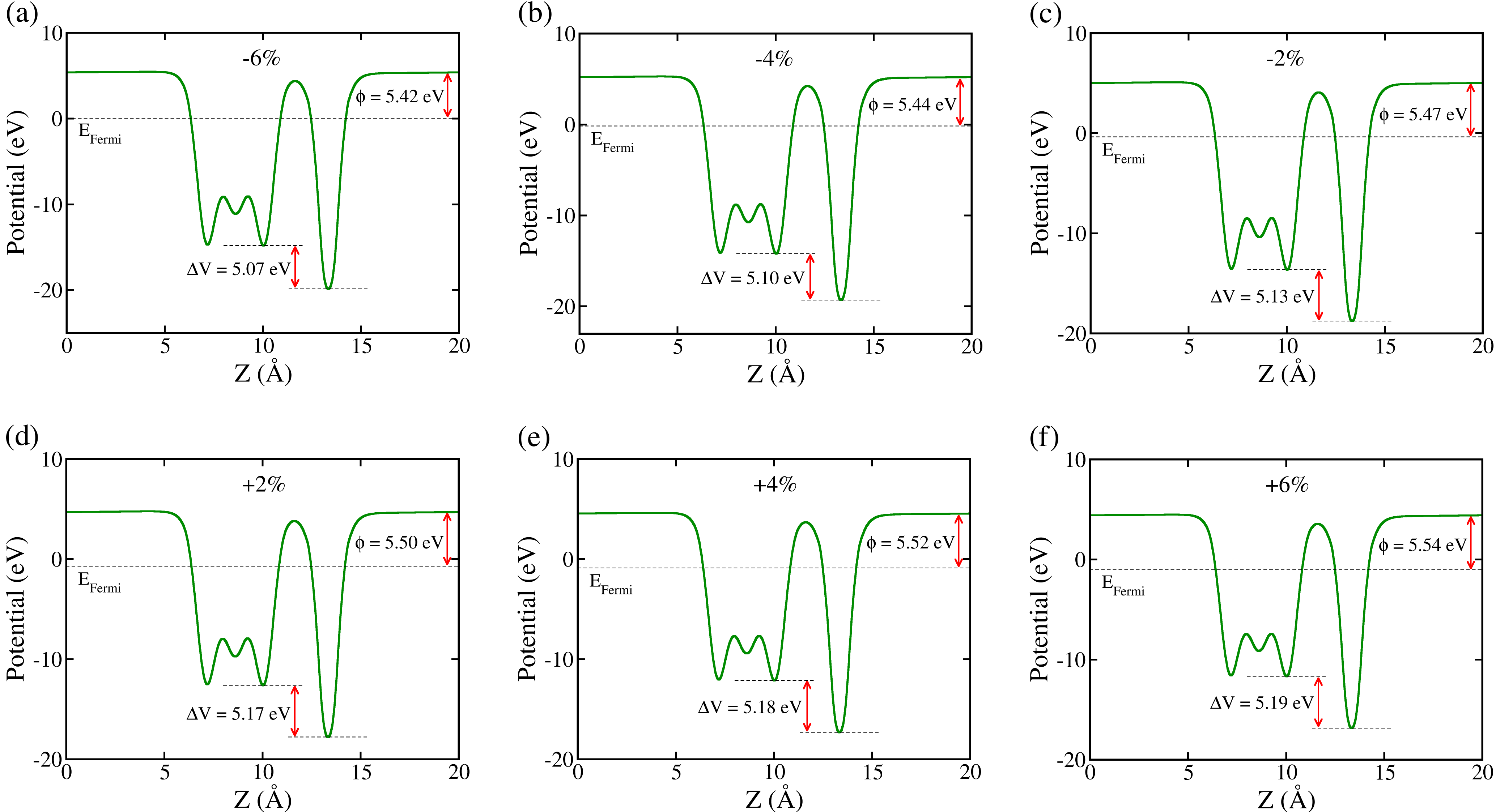}
\end{overpic}
\caption{Average electrostatic potentials of the GeC/ReS$_2$ heterostructure under biaxial strain ranging from -6\% to +6\%: (a) -6\%, (b) -4\%, (c) -2\%, (d) +2\%, (e) +4\%, and (f) +6\%.}
\label{fig:potentials-strain}
\end{figure*}

\begin{table*}[ht!]
\centering
\caption{\label{tab:STH}
Solar-to-hydrogen (STH) efficiency ($\eta_{\text{STH}}$, in \%) of the 
GeC/ReS\textsubscript{2} heterostructure at different pH values under biaxial 
strain from $-6\%$ (compressive) to $+6\%$ (tensile). Values repeated down a 
column indicate that $\eta_{\text{STH}}$ is constant over that pH range for the 
given strain. The light absorption efficiency ($\eta_{\text{abs}}$) depends 
only on the band gap and hence only on strain, whereas the carrier usage 
efficiency ($\eta_{\text{cu}}$) and consequently $\eta_{\text{STH}}$ are 
pH-dependent through the overpotentials $\chi(\mathrm{H_2})$ and 
$\chi(\mathrm{O_2})$. All values are calculated at $T=298.15$~K using the 
AM1.5G solar spectrum.}
\setlength{\tabcolsep}{10pt}
\normalsize
\begin{tabular}{c|rrrrrrrr}
\hline
\hline
\multicolumn{1}{c|}{} & \multicolumn{7}{c}{Strain} \\
pH & $-6\%$ & $-4\%$ & $-2\%$ & $0$ & $+2\%$ & $+4\%$ & $+6\%$ \\
\hline
\rule{0pt}{11pt}$0$     & $11.20$ & $18.48$ & $28.36$ & $38.02$ & $45.14$ & $54.24$ & $60.68$ \\
$1$     & $12.40$ & $20.22$ & $30.48$ & $41.00$ & $46.78$ & $54.24$ & $60.68$ \\
$2$     & $13.66$ & $22.06$ & $33.00$ & $41.40$ & $46.78$ & $54.24$ & $60.68$ \\
$3$     & $15.04$ & $24.06$ & $34.42$ & $41.40$ & $46.78$ & $54.24$ & $60.68$ \\
$4$     & $16.48$ & $26.12$ & $34.42$ & $41.40$ & $46.78$ & $54.24$ & $60.68$ \\
$5$     & $18.08$ & $27.96$ & $34.42$ & $41.40$ & $46.78$ & $54.24$ & $60.68$ \\
$6$     & $19.76$ & $27.96$ & $34.42$ & $41.40$ & $46.78$ & $54.24$ & $60.68$ \\
$7$     & $21.56$ & $27.96$ & $34.42$ & $41.40$ & $46.78$ & $54.24$ & $60.68$ \\
$8$--$14$ & $22.46$ & $27.96$ & $34.42$ & $41.40$ & $46.78$ & $54.24$ & $60.68$ \\
\hline
\hline
\end{tabular}
\end{table*}

\end{document}